\documentclass[sigconf]{acmart}

\usepackage{enumitem}

\usepackage{soul}
\usepackage{tabularx}
\usepackage{subcaption}

\usepackage{amsmath}
\definecolor{irulerPurple}{RGB}{153, 61, 168}
\definecolor{irulerBlue}{RGB}{24, 115, 205} 

\usepackage{makecell}

\newcommand{\name}{iRULER}
\soulregister\cite7
\soulregister\ref7
\soulregister\name7 
\soulregister\todo7 
\soulregister\toreview7 
\soulregister\rev7 
\soulregister\textcolor7

\newcommand{\hlc}[2][yellow]{{\sethlcolor{#1}\hl{#2}}}
\definecolor{revision_color}{HTML}{FFFFFF}
\newcommand{\rev}[1]{\hlc[revision_color]{#1}} 
\newcommand{\toreview}[1]{\hlc[yellow]{#1}} 

\newcommand{\todo}[1]{\textcolor{Red}{[TODO: #1]}}
\usepackage{pifont}

\newlist{designguideline}{enumerate}{1}
\setlist[designguideline,1]{
  label         = DG\arabic*.,
  leftmargin    = 2.6em,                   
  labelsep      = 0.5em,                  
  itemsep       = 0.0\baselineskip     
}
\newlist{enumerate_researchquestions}{enumerate}{1}
\setlist[enumerate_researchquestions,1]{
  label         = RQ\arabic*.,
  leftmargin    = 2.6em,                   
  labelsep      = 0.5em,                  
  itemsep       = 0.0\baselineskip     
}
\newlist{enumerate_hypotheses}{enumerate}{1}
\setlist[enumerate_hypotheses,1]{
  label         = H\arabic*.,
  leftmargin    = 2.4em,                   
  labelsep      = 0.5em,                  
  itemsep       = 0.0\baselineskip     
}

\AtBeginDocument{%
  }
    
\copyrightyear{2026}
\acmYear{2026}
\setcopyright{cc}
\setcctype{by}
\acmConference[CHI '26]{Proceedings of the 2026 CHI Conference on Human Factors in Computing Systems}{April 13--17, 2026}{Barcelona, Spain}
\acmBooktitle{Proceedings of the 2026 CHI Conference on Human Factors in Computing Systems (CHI '26), April 13--17, 2026, Barcelona, Spain}
\acmPrice{}
\acmDOI{10.1145/3772318.3790539}
\acmISBN{979-8-4007-2278-3/2026/04}

\AtBeginMaketitle{%
\let\oldAtBeginDocument\AtBeginDocument%
\renewcommand\AtBeginDocument[1]{#1}
\setcopyright{cc}
\setcctype{by}
\copyrightyear{2026}
\acmYear{2026}
\acmDOI{10.1145/3772318.3790539}
\acmISBN{979-8-4007-2278-3/26/04}
\acmConference[CHI '26]{Proceedings of the 2026 CHI Conference on Human Factors in Computing Systems}{April 13--17, 2026}{Barcelona, Spain}
\acmBooktitle{Proceedings of the 2026 CHI Conference on Human Factors in Computing Systems (CHI '26), April 13--17, 2026, Barcelona, Spain}
\let\AtBeginDocument\oldAtBeginDocument%
}

\begin{document}
\title{iRULER: Intelligible Rubric-Based User-Defined LLM Evaluation for Revision}

\author{Jingwen Bai}
\orcid{0000-0001-5118-993X}
\affiliation{
\department{Department of Computer Science}\institution{National University of Singapore}
\city{Singapore}
\country{Singapore}}
\email{jingwenbai@u.nus.edu}

\author{Wei Soon Cheong}
\orcid{0009-0001-2179-3455}
\affiliation{
\department{Department of Computer Science}\institution{National University of Singapore}
\city{Singapore}
\country{Singapore}}
\email{weisoon.cheong@u.nus.edu}

\author{Philippe Muller}
\orcid{0000-0002-6765-4020}
\affiliation{
\department{IRIT}\institution{University of Toulouse}
\city{Toulouse}
\country{France}}
\email{philippe.muller@irit.fr}

\author{Brian Y Lim}
\authornote{Corresponding author}
\orcid{0000-0002-0543-2414}
\affiliation{
\department{Department of Computer Science}\institution{National University of Singapore}
\city{Singapore}
\country{Singapore}}
\email{brianlim@nus.edu.sg}

\renewcommand{\shortauthors}{Bai, Cheong, Muller, and Lim}
\begin{abstract}

Large Language Models (LLMs) have become indispensable for evaluating writing. However, text feedback they provide is often unintelligible, generic, and not specific to user criteria. Inspired by structured rubrics in education and intelligible AI explanations, we propose iRULER following identified design guidelines to \textit{scaffold} the review process by \textit{specific} criteria, providing \textit{justification} for score selection, and offering \textit{actionable} revisions to target different quality levels. To \textit{qualify} user-defined criteria, we recursively used iRULER with a rubric-of-rubrics to iteratively \textit{refine} rubrics. In controlled experiments on writing revision and rubric creation, iRULER most improved validated LLM-judged review scores and was perceived as most helpful and aligned compared to read-only rubric and text-based LLM feedback. Qualitative findings further support how iRULER satisfies the design guidelines for user-defined feedback. This work contributes interactive rubric tools for intelligible LLM-based review and revision of writing, and user-defined rubric creation.
\end{abstract}


\begin{teaserfigure}
  \centering
  \includegraphics[width = 1.00\linewidth]{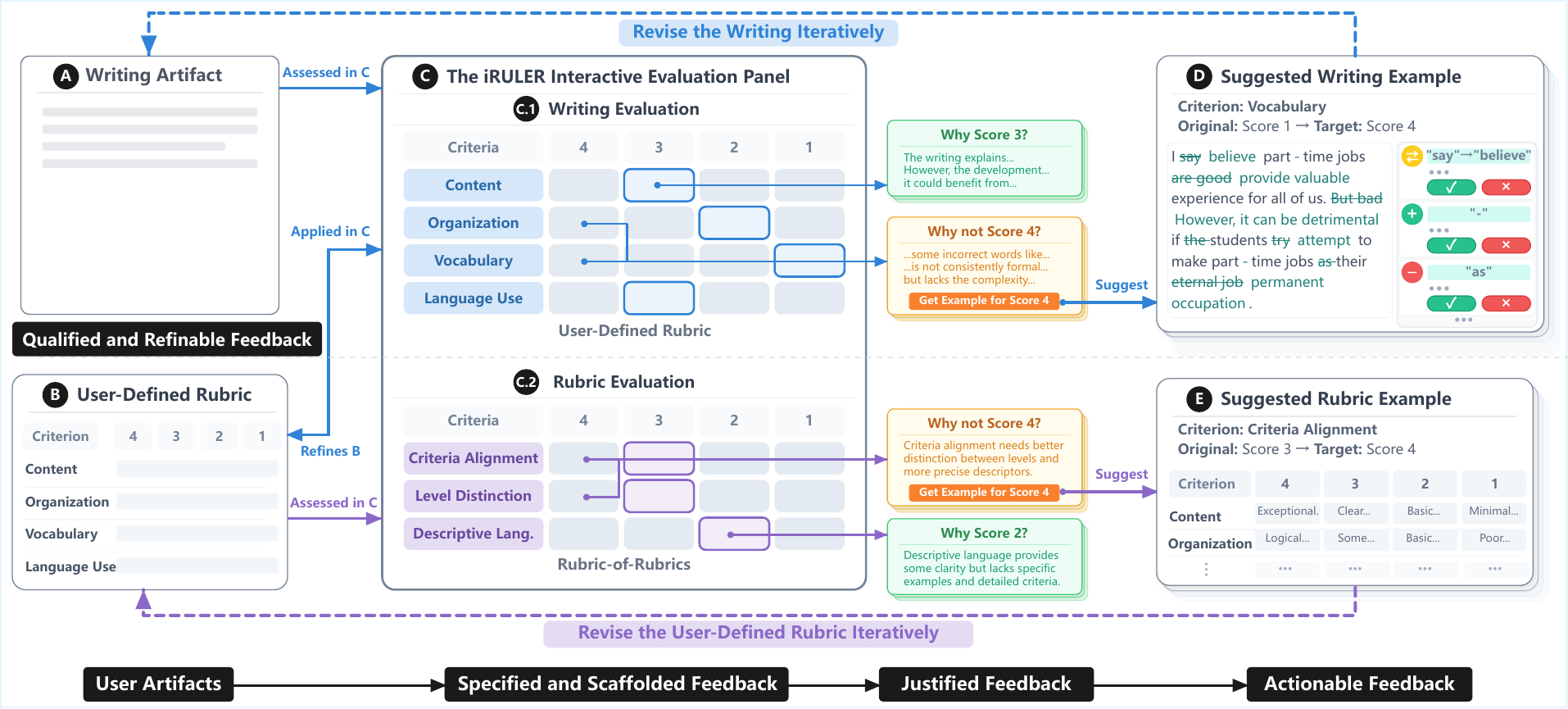}
  \caption[]{
  The \name{} Dual-Level Feedback Workflow. 
  \name{} recursively applies the same feedback mechanism at two levels. 
  Top: The (A) Writing is evaluated against (C.1) a User-Defined Rubric to suggest (D) revisions via \textit{Specific}, \textit{Scaffolded}, \textit{Justified}, and \textit{Actionable} feedback. 
  Bottom: The (B) User-Defined Rubric itself is evaluated against (C.2) a Rubric-of-rubrics to ensure it is \textit{Qualified}. 
  A cross-level Refinement Loop facilitates iteration, keeping the rubric \textit{Refinable} and aligned with user goals.
  }
  \Description{A conceptual diagram illustrating the iRULER system's recursive workflow, aligned with six design guidelines (Specific, Scaffolded, Justified, Actionable, Qualified, Refinable). 
The diagram is split into two interconnected loops. 
The top loop shows a Writing Artifact being evaluated by a User-Defined Rubric, generating justified explanations and actionable revision suggestions. 
The bottom loop recursively evaluates the User-Defined Rubric itself against a meta-rubric (Rubric-of-Rubrics) to ensure quality. 
Crucially, a prominent feedback arrow connects the actionable writing feedback back to the rubric definition stage, visualizing the 'Refinable' guideline where users update their criteria based on concrete revision examples.}
  \label{fig:teaser}
\end{teaserfigure}

\maketitle
\section{Introduction}

Recent advances in Large Language Models (LLMs) have introduced a paradigm shift in automated feedback, promising to provide fast, personalized guidance at scale~\cite{Steiss2024Comparing, laban2024beyond}. Moving beyond the surface-level corrections of classic Automated Writing Evaluation systems, LLMs can generate sophisticated, content-aware evaluation that aligns with human graders on many evaluative dimensions~\cite{chiang2023can, escalante2023ai, chang2021exploring}. Writers often find this AI-generated feedback more detailed and understandable than traditional instructor comments~\cite{Meyer2023Using}, positioning LLMs as powerful ``personal writing tutors''~\cite{rad2024using, banihashem2024feedback}.
However, the utility of LLMs for evaluative feedback is constrained by several challenges. 
Their opaque reasoning processes~\cite{stechly2023gpt} make it difficult for users to interpret the justification behind a given score. 
This opacity is compounded by potential hallucinations~\cite{huang2025survey} and misalignment with human judgment~\cite{silva2024Analysis}, making feedback difficult to trust or act upon~\cite{jansen2024empirische}.
This creates a critical gap: writers have access to instantaneous feedback, but it is not consistently structured or actionable enough to guide meaningful revision.
Previous research on eXplainable AI~(XAI) demonstrates that transparency in AI decision-making boosts user understanding and skill development~\cite{lim2009and, lai2023selective}. 
This principle is vital in learning contexts, where effective feedback must not only evaluate, but also explain evaluations and guide subsequent actions~\cite{Lipnevich2022Unraveling, poulos2008effectiveness}. 

Inspired by the use of rubrics in education to provide structured feedback~\cite{panadero2013rubrics}, we propose a new approach to scaffold LLM-generated evaluations. A rubric is a scoring tool that lists criteria for evaluating work alongside descriptions of different levels of performance quality~\cite{brookhart2015quality}. By clarifying what counts as quality work, well-crafted rubrics enable focused, transparent feedback and help writers internalize standards and self-regulate~\cite{panadero2013rubrics, jonsson2007use, stevens2023introduction}; rubrics also can be a way to explain judgment~\cite{matsuyama2023iris, stevens2023introduction}; empirical studies further show that rubric-based feedback yields greater improvements than generic comments~\cite{hasan2022effect, wolf2007role, reddy2010review}. Furthermore, involving students in rubric co-creation leads to better learning experiences and outcomes~\cite{yan2024rubric, er2021collaborative}. In essence, well-defined rubrics serve as a common guide, helping writers and evaluators to develop a shared understanding of quality.
Recent works, such as LLM-Rubric~\cite{hashemi2024llm} and Prometheus~\cite{kim2023prometheus}, 
show that LLMs can score text against rubrics,
but neglect to provide criterion-specific explanations or actionable guidance for collaborative revision,
and were designed mainly for automated benchmarking with static, predefined rubrics.

However, the efficacy of rubric-based feedback hinges on the quality of the rubric itself. Creating and qualifying a rubric is a non-trivial task~\cite{brookhart2013create}.
Writers must translate genre knowledge into specific criteria and calibrate performance levels for scaffolding, which requires expertise~\cite{jonsson2007use,reddy2010review,stevens2023introduction}. 
Without this expertise, many users, including students, novice instructors, and content editors, create unqualified criteria and poorly scaffolded levels,
resulting in ambiguity, lower inter-rater reliability, and weaker feedback guidance~\cite{jonsson2007use,panadero2013rubrics}. 
This expertise barrier often results in the adoption of generic, \emph{one-size-fits-all} rubrics that overlook specific task goals or genre nuances~\cite{brookhart2015quality,panadero2013rubrics}. 
Moreover, recent work shows that evaluation criteria tend to drift as evaluators see more outputs, leading them to add, split, and revise criteria~\cite{shankar2024validates}.
Rubrics therefore should be treated as evolving rather than static templates.



To address these challenges, 
we draw on works on rubrics as structured feedback tools in education~\cite{wolf2007role,reddy2010review} and to human-interpretable explanations in XAI~\cite{lim2009and,ehsan2021expanding,liao2020questioning}
to derive six design guidelines for user-defined AI feedback.
Feedback should be \textit{specific}, \textit{scaffolded}, \textit{justified}, \textit{actionable}, \textit{qualified}, and \textit{refinable}.
Taken together, these guidelines call for structuring LLM feedback into \textit{specific} criteria with \textit{scaffolded} levels, providing \textit{justified} criteria-level rationales, and offering \textit{actionable} revision suggestions,
while also requiring that user-defined criteria themselves be \textit{qualified} and iteratively \textit{refinable} through a rubric-of-rubrics~\cite{arter2000rubrics,arter2001scoring,stevens2023introduction}.



To embody these guidelines, 
we present \textbf{\name{}}, an interactive system that provides \textbf{I}ntelligible \textbf{R}ubric-based, \textbf{U}ser-defined \textbf{L}LM \textbf{E}valuation for \textbf{R}evision. 
\name{} functions through two distinct but integrated applications. 
First, a \textbf{writing revision} application uses a qualified rubric to drive \textit{specific}, \textit{scaffolded}, \textit{justified} and \textit{actionable} feedback in guiding users' revisions. 
Second, a \textbf{rubric creation} application recursively applies these four design guidelines to the rubric itself;
a \textit{rubric-of-rubrics}~\cite{arter2000rubrics} provides the same principled feedback to help users \textit{qualify} their criteria. 
By using these two applications end-to-end, \name{} ensures that the user-defined rubric is \textit{refinable} as it is applied to the artifacts.



In two between-subjects experiments, a \textbf{writing revision experiment} (\textit{N} $=$ 48) and a \textbf{rubric creation experiment} (\textit{N} $=$ 36), we found that
\name{} not only improved the quality of both the writing and user-designed rubrics, but was also perceived by users as more helpful, aligned with their judgment, and controllable compared to baseline systems, and reported higher confidence gains. 
Although \name{} is primarily designed as a collaborative tool rather than a tutoring tool, our study nevertheless suggests that interacting with \name{} may support transferable writing skills.
This aligns with work showing that rubrics can improve learning outcomes~\cite{wolf2007role, reddy2010review} and that transparent, explanation-based AI feedback enables ``forward simulation'' of reasoning~\cite{hase2020evaluating, bo2024incremental}.
Our \textbf{end-to-end experiment} (\textit{N} $=$ 6) further demonstrates that \name{}
supports user-driven rubric refinement with diverse, self-chosen writing genres.
Qualitative analyses indicate that \name{} satisfies our design guidelines for user-defined feedback, scaffolds user agency rather than over-reliance, and supports human-AI co-evaluation.

Since we had used LLM-as-a-judge for feedback scoring and scalable evaluation, we validated our LLM-based evaluation pipeline on an expert-annotated dataset and in a \textbf{human expert validation experiment} (\textit{N} $=$ 2), finding that our system's scores align well with human expert judgments.


Our \textbf{contributions} are: 


\begin{enumerate}
    \item 
    \textbf{Design Guidelines for User-Defined Feedback.}
    We derive six design guidelines---Specific, Scaffolded, Justified, Actionable, Qualified and Refinable---for user-defined rubrics.
    We articulate a rubric-based framework to serve as structured scaffolds enabling transparent and actionable feedback in LLM-based systems.


    \item \textbf{\name{}}, 
    an interactive, LLM-infused feedback tool supporting user-defined rubrics.
    We applied it to two applications based on rubrics---\emph{Writing Revision} and recursively, \emph{Rubric Creation}.
    These applications can be used separately by two stakeholders (e.g., student, teacher), or end-to-end by a single user to iteratively apply and refine feedback criteria.
    Thus, iRULER advances beyond prior structured feedback by enabling the recursive co-creation and refinement of the evaluation criteria itself.
    
\end{enumerate}


We provide empirical evidence that \name{} produces higher-quality artifacts 
and is perceived by users as more effective than generic, read-only feedback. 
Alongside empirical evidence validating the effectiveness of our approach, we discuss insights, generalizations, limitations, and future work.

\section{Related Work}
We examine the current gaps in LLM-based feedback, need for and usage of rubrics, and opportunity of using explainable AI to enhance AI-based feedback.

\subsection{LLM-Powered Feedback} 
Large Language Models (LLMs) present an unprecedented ability to facilitate human-AI collaboration in various domains including creative tasks~\cite{rezwana2023designing, ippolito2022creative}, writing~\cite{zhang2023visar, laban2024beyond, lee2024design}, programming~\cite{wu2021ai, leung2023best, yen2024coladder} to even evaluation~\cite{ashktorab2024aligning, shankar2024validates, sun2024reviewflow}. 
This collaborative stance can enhance users' sense of agency as they are actively involved in decision-making rather than passively receiving answers~\cite{westphal2023decision}.
In writing, LLMs have enabled a new generation of feedback tools capable of generating sophisticated commentary, moving beyond the surface-level focus of classic Automated Writing Evaluation systems~\cite{ding2024automated}. 
Systems like LLM-Rubric~\cite{hashemi2024llm} have shown the feasibility of prompting LLMs to perform multi-dimensional evaluations against a given rubric, promising to align AI scoring with human values~\cite{escalante2023ai, kim2023prometheus, lin2024wildbench}.

To explain these evaluations, many systems rely on Chain-of-Thought (CoT) prompting, where the model generates a rationale before producing a score~\cite{kim2023prometheus, liu2023g}.
However, this approach presents interactional and reliability challenges.
First, CoT feedback is typically delivered as an unstructured stream of text within a chat interface. 
While expressive, this lack of scaffolding makes it difficult for users to interactively interrogate specific criteria or justify the reasoning behind a specific score~\cite{wang2023mint}.
Second, the generated explanations can be plausible-sounding post-hoc justifications that may be unfaithful to the model's actual decision parameters~\cite{turpin2023language}.
Consequently, even when guided by criteria, 
users are often left with untrustworthy and unactionable feedback~\cite{tang2024harnessing, gandolfi2025gpt}. 
This creates a critical gap: we have a powerful generative engine (the LLM) but lack a transparent, structured mechanism to reliably steer its behavior and make its evaluative reasoning understandable to the user~\cite{kim2025fostering}. This underscores the importance of moving beyond issue detection towards interactive systems that support user sense-making and action, as seen in systems like \emph{Friction}~\cite{zhang2025friction} and \emph{InkSync}~\cite{laban2024beyond}.


\subsection{Rubrics for Structured Evaluation}
To address the need for a ``transparent, structured mechanism''~\cite{kim2025fostering} we turn to rubrics~\cite{jonsson2014rubrics, bearman2021can}. Rubrics are widely used in education as structured scoring tools, especially for tasks like writing~\cite{panadero2013rubrics, jonsson2007use, hasan2022effect}, which provide an explicit basis for assessment by defining clear criteria and performance levels~\cite{hasan2022effect, Lipnevich2022Unraveling}. Research shows their potential is maximized not as static grading tools, but as interactive scaffolds that guide learning and revision~\cite{stevens2023introduction, wolf2007role}.

However, traditional rubric-based feedback often fails to realize this interactive potential~\cite{Ene2016Rubrics}, providing scores with minimal explanation and little actionable insight~\cite{yune2018holistic, wollenschlager2016makes, klein1998analytic}.
Studies on rubric co-creation demonstrate that giving users ownership over criteria significantly enhances agency and feedback uptake~\cite{yan2024rubric, er2021collaborative}. Concurrently, systems like \emph{EvalLM} and \emph{EvalAssist} treat evaluation criteria as iteratively refinable, user-defined artifacts~\cite{kim2024evallm, ashktorab2024aligning}. 
This body of work establishes a key premise: an ideal evaluative tool should be built upon a rubric that is both interactive and user-driven~\cite{afsar2021assessing, madaan2023self}. 
While these systems provide the necessary structure, they stop at judgment: the per-criterion scores they generate seldom include in-depth justification, and fail to translate into actionable revision guidance. 
Assessment research has also proposed \emph{rubrics for rubrics} that evaluate the quality of assessment criteria themselves~\cite{arter2001scoring, arter2000rubrics, stevens2023introduction}, yet in practice the qualification and refinement of user-defined rubrics is often left to users' intuitions. 
Therefore, the question of how to make the AI's application of that structure transparent remains.



\subsection{Explainable AI for Evaluative Feedback}
The challenge of making rubric-based reasoning intelligible leads us to the goal of making AI systems understandable to their users~\cite{abdul2018trends, capel2023human, xu2023transitioning}. 
Seminal work in this area has reframed explanation from a static output to an interactive dialogue~\cite{lai2023selective, liao2020questioning, wang2022interpretable, wang2023gam}.

Previous research demonstrated that enabling users to ask fundamental questions, such as ``Why?'', ``Why not?'', and ``How to?'', can significantly improve their mental model and trust in the system~\cite{lim2009and, lim2009assessing}. Subsequent work in Explainable AI (XAI) has consistently affirmed the value of these contrastive (``Why not...?'') and counterfactual (``How to...?'') explanations for making AI reasoning more meaningful~\cite{zhang2022towards, ehsan2019automated}. This body of research shows that users leverage such explanations not merely to understand a single output, but to calibrate trust, improve skills, and provide more effective inputs in future interactions~\cite{lee2023understanding, ahn2024impact, liu2023increasing}. This aligns with recent calls for human-centered XAI to design explanation systems tailored to the specific needs of end-users in their context~\cite{kim2023help, panigutti2023co, morrison2024impact}. Recent research like IRIS~\cite{matsuyama2023iris} shows that integrating domain rubrics into AI models not only improves performance but also addresses critical concerns about transparency and trust, which is valuable for justifying subjective judgment.
While prior studies have identified effective explanations, a key challenge remains: how to systematically implement them in complex areas providing feedback on writing and designing user-defined criteria~\cite{fan2022human, gayed2022exploring, zhang2025friction}.

\section{Design Guidelines for User-defined Feedback}
\label{sec:design_guidelines}
Based on the literature review and reflections, we derive six design guidelines (DGs) for user-defined feedback, 
that feedback needs to be: specific, scaffolded, justified, actionable, qualified, and refinable.

\begin{enumerate}
    \item[{DG1.}] \textbf{Specific.} 
    Feedback should be specified in terms of explicit criteria~\cite{ngoon2018interactive, kim2024evallm} that users can judge for themselves and take remedial action~\cite{wadhwa2025evalagent}; 
    generic feedback of vague or inconsistent concepts is not helpfully understandable and  actionable~\cite{meyer2024using, khraisha2024can, shin2025visualizationary}. 

   \item[{DG2.}] \textbf{Scaffolded.} 
    Defining evaluation criteria is challenging~\cite{davidson2005evaluation, reddy2010review}. Not only do users need to decide what is important, they have to determine what passes or fails each criteria.
    Rather than leave this open-ended, which is cognitively demanding~\cite{flower1980cognition}, feedback criteria can be scaffolded via a structured template~\cite{gielen2015structuring, ajjawi2022feedback}.
    Rubrics satisfy this by defining levels for each criteria, and supports weighting criteria by importance all in a tabular layout~\cite{jonsson2007use,stevens2023introduction, brookhart2015quality}.

    \item[{DG3.}] \textbf{Justified.} 
    Rather than providing feedback based on opaque, subjective opinion, feedback needs to be justified with rationale and explanation~\cite{ngoon2018interactive}.
    With AI generated judgment, this can be explained with explainable AI (XAI) methods. 
    Rubrics are one way to justify the judgment score~\cite{matsuyama2023iris, kumar2020explainable}.
    Yet, the judgment of each criteria may be unintelligible~\cite{lim2009and}, thus deeper levels should be further justified too~\cite{lim2011design, xie2024grade} to explain \textit{why} specific criteria level were chosen, and \textit{why not} other levels.

   \item[{DG4.}] \textbf{Actionable.} 
    A primary goal of feedback is to improve ones work~\cite{wadhwa2025evalagent}. Feedback needs to be actionable for users to make effective changes~\cite{ngoon2018interactive, mothilal2020explaining}.
    Therefore, suggestions should be provided via counterfactual feedback of \textit{how to} achieve desired target outcomes~\cite{lyu2024if, wang2022interpretable}.

    \item[{DG5.}] \textbf{Qualified.} 
    Although any feedback is appreciated, they may be erroneous or unreliable, especially when developed by users who are not domain or topic experts~\cite{lucassen2013topic, brand2017source}. 
    Hence, it is important to judge the feedback, and even provide feedback on the feedback~\cite{winstone2017supporting, hattie2007power}.
    Similarly, this feedback should be \textit{specific}, \textit{scaffolded}, \textit{justified} and \textit{actionable}.
    Thus, we propose a recursive use of rubric-based feedback on the rubric-feedback using the \textit{rubric-of-rubrics}~\cite{arter2000rubrics, arter2001scoring, stevens2023introduction}.

    \item[{DG6.}] \textbf{Refinable}.
    Moreover, despite accommodating user-defined criteria, users may change their mind (i.e., criteria drift~\cite{shankar2024validates} or evolution~\cite{szymanski2024comparing}) and iteratively refine feedback criteria~\cite{kim2024evallm, ashktorab2025evalassist, gebreegziabher2025metricmate}.
    Hence, user-defined rubrics should be refinable through the end-to-end loop of rubric editing with feedback, and rubric usage for writing revision.

\end{enumerate}


\section{iRULER System}

\begin{figure*}
    \centering
    \includegraphics[width = 0.92\linewidth]{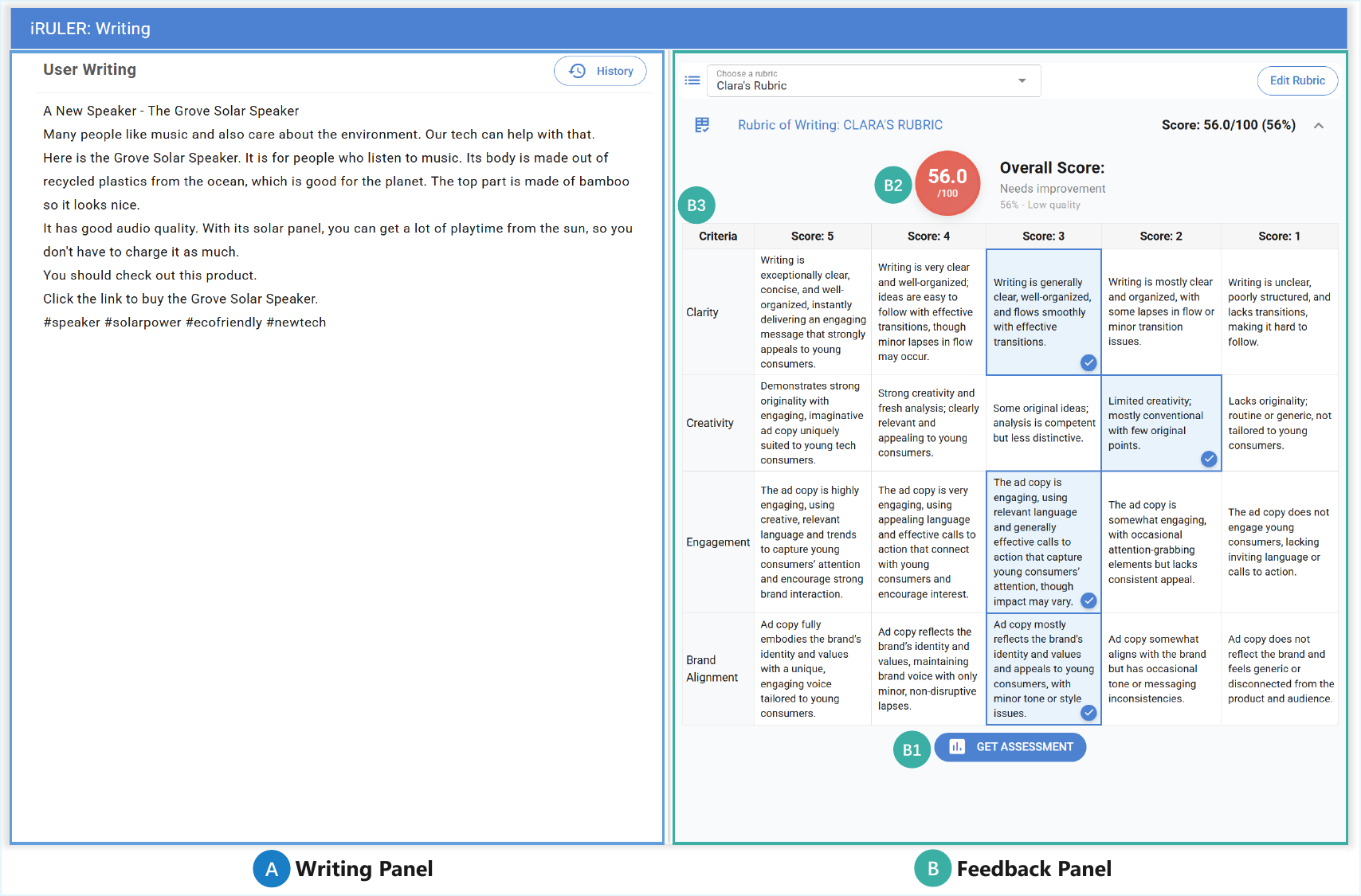}
    \caption[]{
    The \name{} Writing Revision User Interface. 
    The (A) Writing Panel displays the current draft. 
    The (B) Feedback Panel provides interactive evaluation, allowing users to B1) trigger assessment, B2) see the overall score, B3) interact with the rubric.
  }
    \Description{The user interface consists of two main panels. The left Writing Panel displays an editor containing a user-written ad copy for a eco-friendly Bluetooth speaker. 
    The right Feedback Panel shows Clara's Rubric in a 4x5 rubric table with an overall score of 56.0/100. Four criteria (rows) are evaluated on a 5-point scale (columns), and individual criterion scores are highlighted in blue: Clarity (3/5), Creativity (2/5), Engagement (3/5) and Brand Alignment (3/5). Each score level includes detailed descriptive text, with higher scores indicating higher quality on the respective criterion. The panel includes ``Edit Rubric'' and ``Get Assessment'' interactive buttons.}
    \label{fig:writing-interface}
\end{figure*}

We propose \name{} to implement the aforementioned design guidelines for user-defined feedback.
iRULER provides feedback that is 
i) \textit{\textbf{specified}} with criteria, 
ii) \textit{\textbf{scaffolded}} using a rubric, 
iii) \textit{\textbf{justified}} with intelligible explanations of scores and subscores, 
iv) \textit{\textbf{actionable}} for users to make changes, and 
v) \textit{\textbf{qualified}} by recursively providing feedback on the rubrics.
It consists of two applications, one for Writing Revision (i--iv), and one for Rubric Creation (i--v).
Using both applications end-to-end can make the feedback 
vi) \textit{\textbf{refinable}} through iterative editing and usage of the rubric.



\subsection{Writing Revision}
\label{sub-sec:writing_feedback_system}

Fig. \ref{fig:writing-interface} shows the Writing Revision User Interface.
The writing application is divided into two main parts:
A) Writing Panel on the left, where users freely type and edit any text, and 
B) Feedback Panel on the right.
The feedback consists of showing the overall score and explanatory rubric.

\subsubsection{Concise Quality Score} 
We implement LLM-as-a-judge~\cite{gu2024survey} to score the writing based on the rubric (See Appendix~\ref{appendix:iruler-evaluation-prompts} for prompt details).
When users click \textit{Get Assessment} (Fig.~\ref{fig:writing-interface} B1), the LLM returns a score for each criterion based on its level descriptions (Fig.~\ref{fig:writing-interface} B3) and the system aggregates these into an overall 0--100 score (Fig.~\ref{fig:writing-interface} B2).

Each criterion is scored on an integer level $1,\dots,L$ ($L$ can range from 3 to 6 in the iRULER UI). 
Let $s_k$ be the selected score for criterion $k$, and $w_k$ its weight in percentage, with the total weight summing to 100. 
The overall score is the weighted sum of normalized criterion scores:
{\[
\text{Overall Score} = (w_1\cdot s_1 + w_2\cdot s_2 +...+w_k\cdot s_k)/{L}.
\]}


For example, there are four criteria with weights $\{25, 20, 30, 25\}$\footnote{In our system, hovering over a criterion name reveals its weight.} and five performance levels ($L = 5$) in Fig.~\ref{fig:writing-interface}. 
Selecting levels $\{3, 2, 3, 3\}$ yields: 
Overall Score $= (25 \cdot 3 + 20 \cdot 2 + 30 \cdot 3 + 25 \cdot 3) / 5 = 56.$



\subsubsection{Rubric} 
This explains how to judge the writing based on \textit{criteria} (Fig. \ref{fig:writing-interface} B3, \textbf{DG1: Specific}), in table rows, and with explicit \textit{levels} in columns (\textbf{DG2: Scaffolded}).
This structure helps users to examine each criteria carefully and check how well the writing satisfies each level.
For the case in Fig. \ref{fig:writing-interface}, the writing has middling clarity, engagement and brand alignment (all Score 3), and low creativity (Score 2).
However, it may be unclear how the levels were selected, thus, additional explanations are provided as justification (\textbf{DG3: Justified}).


\begin{figure*}
    \centering
    \includegraphics[width = 1\linewidth]{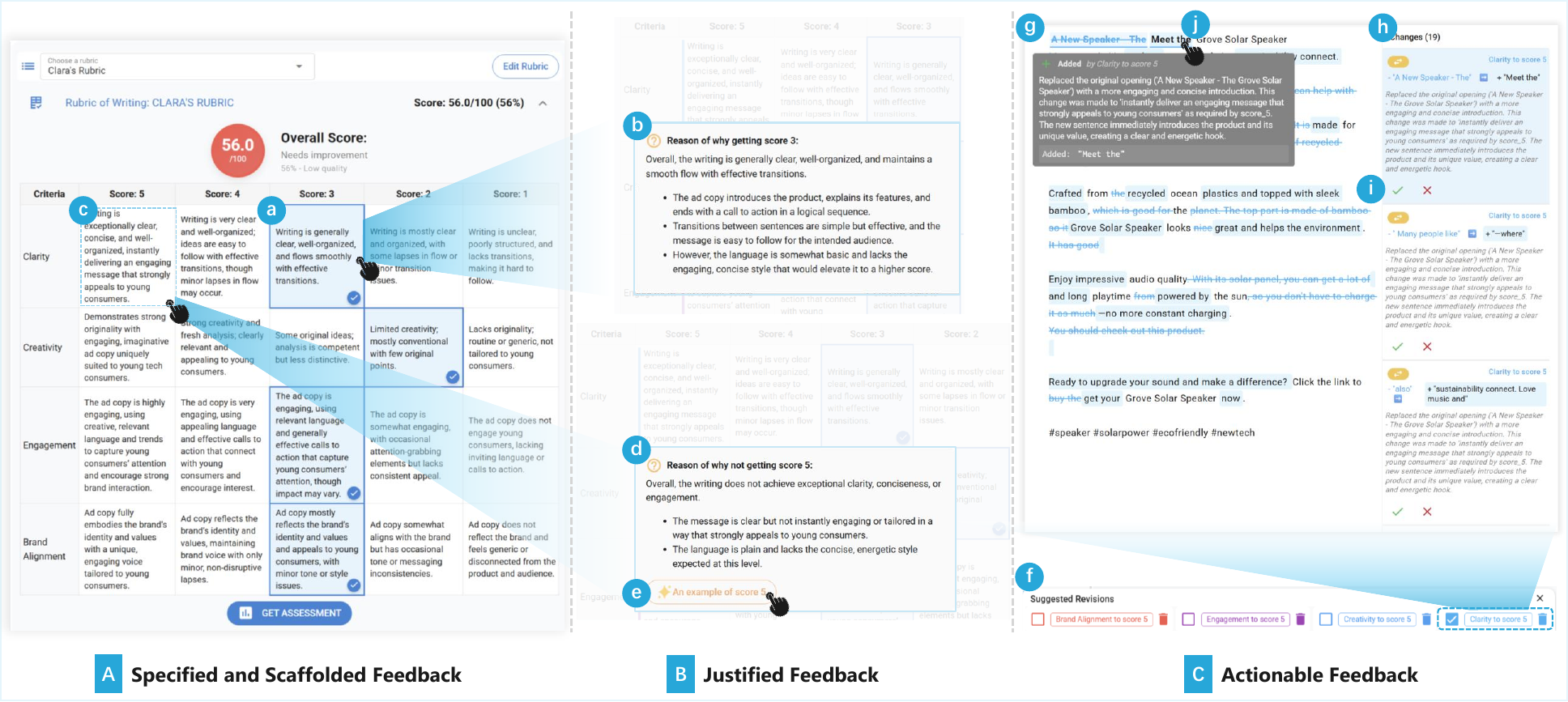}
    \caption[]{The interactive revision workflow in \name{} Writing Revision User Interface. 
    Initial rubric scores provide Specific and Scaffolded feedback. Users can then get Justified feedback by clicking for (b) \textit{Why} and (d) \textit{Why Not} explanations. Requesting (e) a \textit{How To} example generates Actionable, tracked revisions (g-i) that further Scaffold the writing process.}
    \Description{A three-phase diagram illustrates the iRULER writing revision workflow.
    Phase A (left, labeled ``Specified and Scaffolded Feedback'') shows a rubric table. For the ``Clarity'' criterion, score 3 is highlighted with a checkmark. A cursor points at the cell of (Clarity, Score 3), expanding into Phase B's ``Reason of why getting score 3'' pop-up box. Another cursor points at (Clarity, Score 5), expanding into Phase B's ``Reason of why not getting score 5'' pop-up box.
    Phase B (middle, labeled ``Justified Feedback'') displays two pop-up boxes with example text. The top box, titled ``Reason of why getting score 3,'' contains bullet points explaining the score. The bottom box, ``Reason of why not getting score 5,'' explains the deficiencies and includes a button labeled ``An example of score 5.''
    Phase C (right, labeled ``Actionable Feedback'') shows a revised text on the left with color-coded highlights indicating changes. On the right, a series of ``change cards'' corresponds to these edits. Each card details a specific change (e.g., replacing 'A New Speaker - The' with 'Meet the'), provides a text rationale, and has tick and cross buttons for user to accept or decline the edit.}
    \label{fig:writing-flow}
\end{figure*}

\subsubsection{Walkthrough}
We illustrate \name{}'s justified and actionable feedback process through a usage scenario with Leo, a user seeking to improve his writing (Fig. \ref{fig:writing-flow}). 
The workflow begins after an initial assessment scores Leo's writing against the rubric.
To diagnose his performance, Leo first clicks (a) a scored cell, which reveals (b) a \textit{Why} explanation. He then explores (c) a higher-level cell for (d) a \textit{Why Not} explanation (\textbf{DG3: Justified}). 
Building on this understanding, Leo requests (e) a \textit{How To} counterfactual example (\textbf{DG4: Actionable}, see Appendix~\ref{appendix:how-to-writing-prompt} for prompt details). 
\name{} responds by rendering (f) suggested revisions as discrete, (g) color-coded edits directly within Leo's text. Each suggestion is detailed in (h) a ``change card'' that explains the rationale and provides (i) granular Accept/Reject controls\footnote{All writing revision conditions (including baselines) possess this feature for tracking changes.}.

\begin{figure*}
    \centering
    \includegraphics[width = 0.92\linewidth]{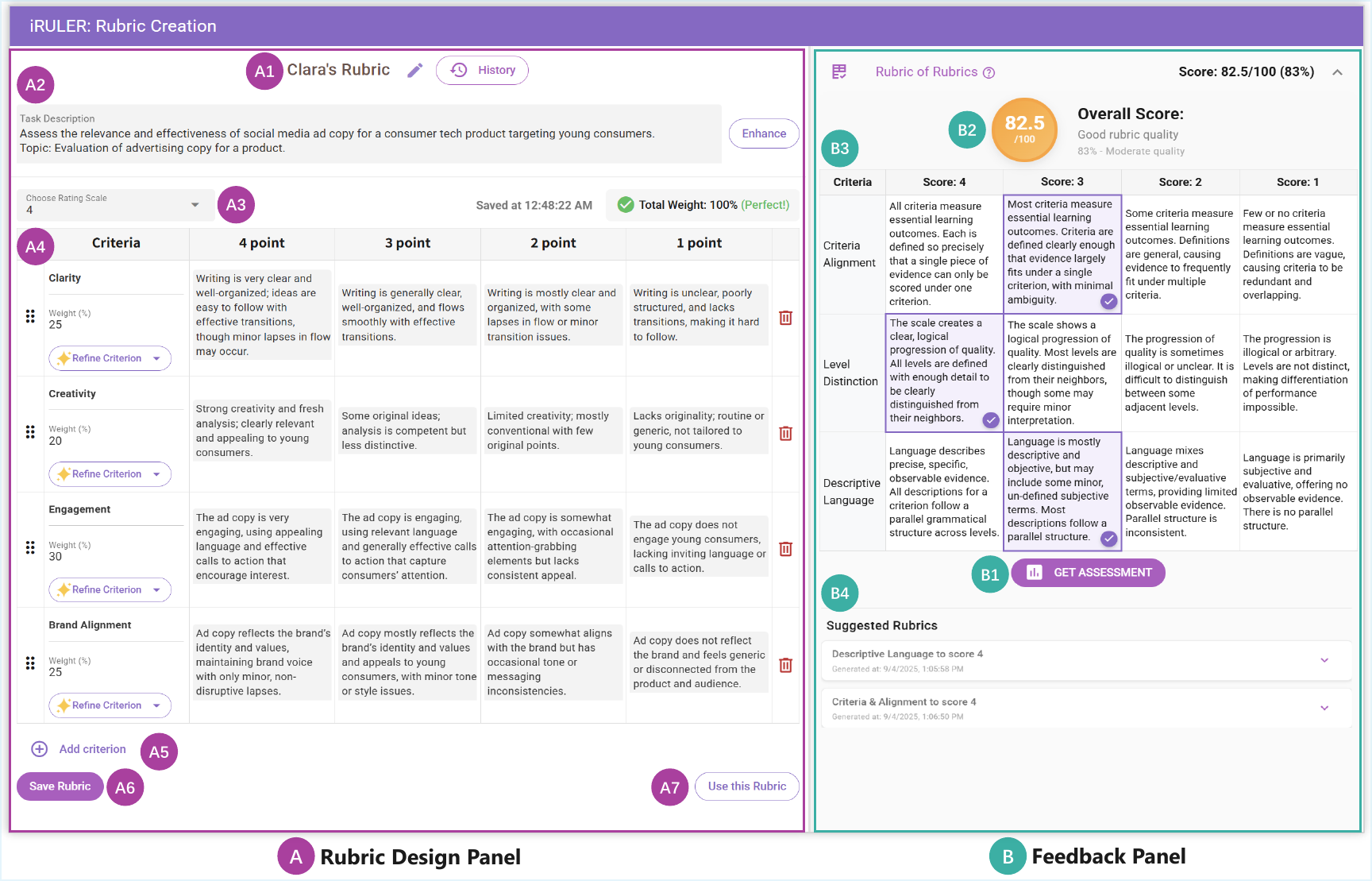}
    \caption[]{The \name{} Rubric Creation User Interface, which allows users to design a rubric in the (A) Design Panel and receive immediate meta-level feedback on its quality in the (B) Feedback Panel. This includes B2) a quality score, B3) a detailed breakdown, and B4) improvement suggestions.}
    \Description{The Rubric Design Panel on the left shows an interactive user interface containing Clara's Rubric, mid-configuration. Above the rubric is a text box for the Task Description, and a drop-down box allowing Clara to choose the Rating Scale, currently at 4 levels. The rubric is displayed in a table format: there are 4 columns, one for each rating level from 4 to 1, and four rows, one for each configurable criteria in Clarity, Creativity, Engagement, and Brand Alignment. Each cell contains configurable descriptive text explaining performance expectations at that level. Additional interactive elements include "Refine Criterion" options for each row, plus "Add criterion" and "Save Rubric" buttons at the bottom.     
    The right Feedback Panel shows the "Rubric of Rubrics" meta-rubric in a 3x4 rubric table with an overall score of 82.5/100. Individual criterion scores are highlighted in purple: Criteria Alignment (3/4), Level Distinction (4/4) and Descriptive Language (3/4). The panel includes a "Get Assessment" interactive button. A "Suggested Rubrics" section below lists generated rubric samples such as "Descriptive Language to score 4" and "Criteria Alignment to score 4".}
  \label{fig:rubric-interface}
\end{figure*}

\subsection{Rubric Creation}
\label{sub-sec:rubric_creation_system}

Fig. \ref{fig:rubric-interface} shows the Rubric Creation User Interface; it helps users design rubrics and receive feedback based on the \textbf{rubric-of-rubrics}~\cite{arter2000rubrics, arter2001scoring} 
(see Table~\ref{tab:rubric-of-rubrics} in Appendix~\ref{appendix-user-study-rubric} for more details).
This meta-rubric was synthesized from literature on the evaluation of rubrics~\cite{brookhart2013create, arter2000rubrics, arter2007creating, mullinix2003rubric}.
Similar to the Writing interface, it has two main parts:
A) the Rubric Design Panel for editing and 
B) the Feedback Panel.

\begin{figure*}[t]
  \centering
  \includegraphics[width = 0.8\linewidth]{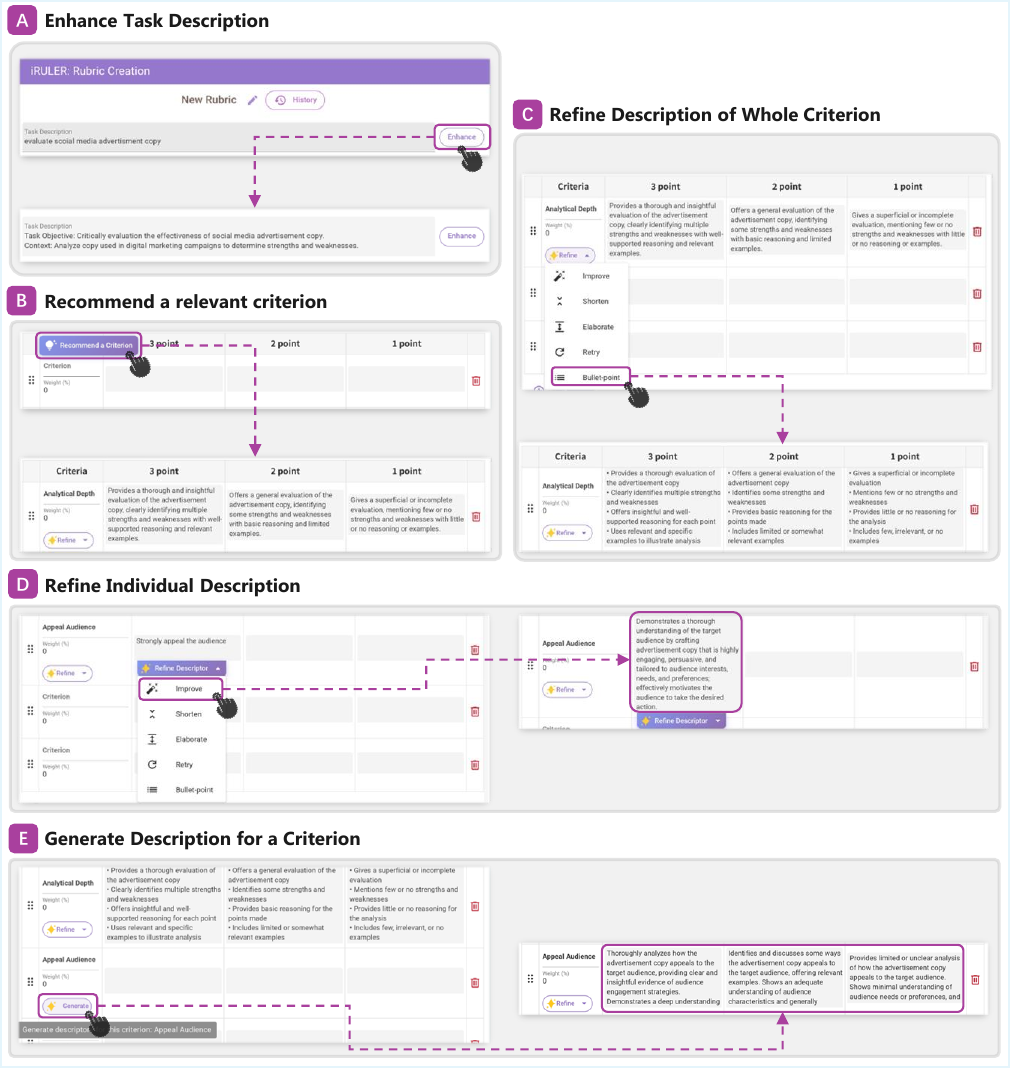}
  \caption[]{The AI-assistive features integrated into the \name{} Rubric Creation UI's Rubric Design Panel. The system supports users throughout the design process by: 
  A) Enhancing the initial task description; 
  B) Recommending relevant criteria based on the description; 
  C) Refining the descriptive text for a whole criterion at once; 
  D) Refining individual performance level descriptors for granular control; and 
  E) Generating a full set of descriptors for a user-defined criterion title.}
  \Description{A diagram showing five AI-assistive features in the Rubric Design Panel of the rubric creation user interface, labeled A through E.
  A) Enhance Task Description: An arrow points from a simple text input field to an enhanced, structured version with "Task Objective" and "Context" sections.
  B) Recommend a relevant criterion: An arrow points from a button labeled "Recommend a Criterion" in an empty rubric row to a fully populated row with the criterion "Analytical Depth" and its level descriptors.
  C) Refine Description of Whole Criterion: An arrow points from a pop-up menu with an option for "Bullet-point" to a version of the "Analytical Depth" criterion where the descriptors are formatted as bulleted lists.
  D) Refine Individual Description: An arrow points from a pop-up menu with an option for "Improve" over a single descriptor cell to a pop-up showing the refined, more detailed text for that specific cell.
  E) Generate Description for a Criterion: An arrow points from a button labeled "Generate" under an empty criterion row named "Appeal Audience" to a version of that row fully populated with new level descriptors.}
  \label{fig:ai-assist-features}
\end{figure*}

\subsubsection{Rubric Design Panel}

\begin{figure*}
    \centering
    \includegraphics[width = 1\linewidth]{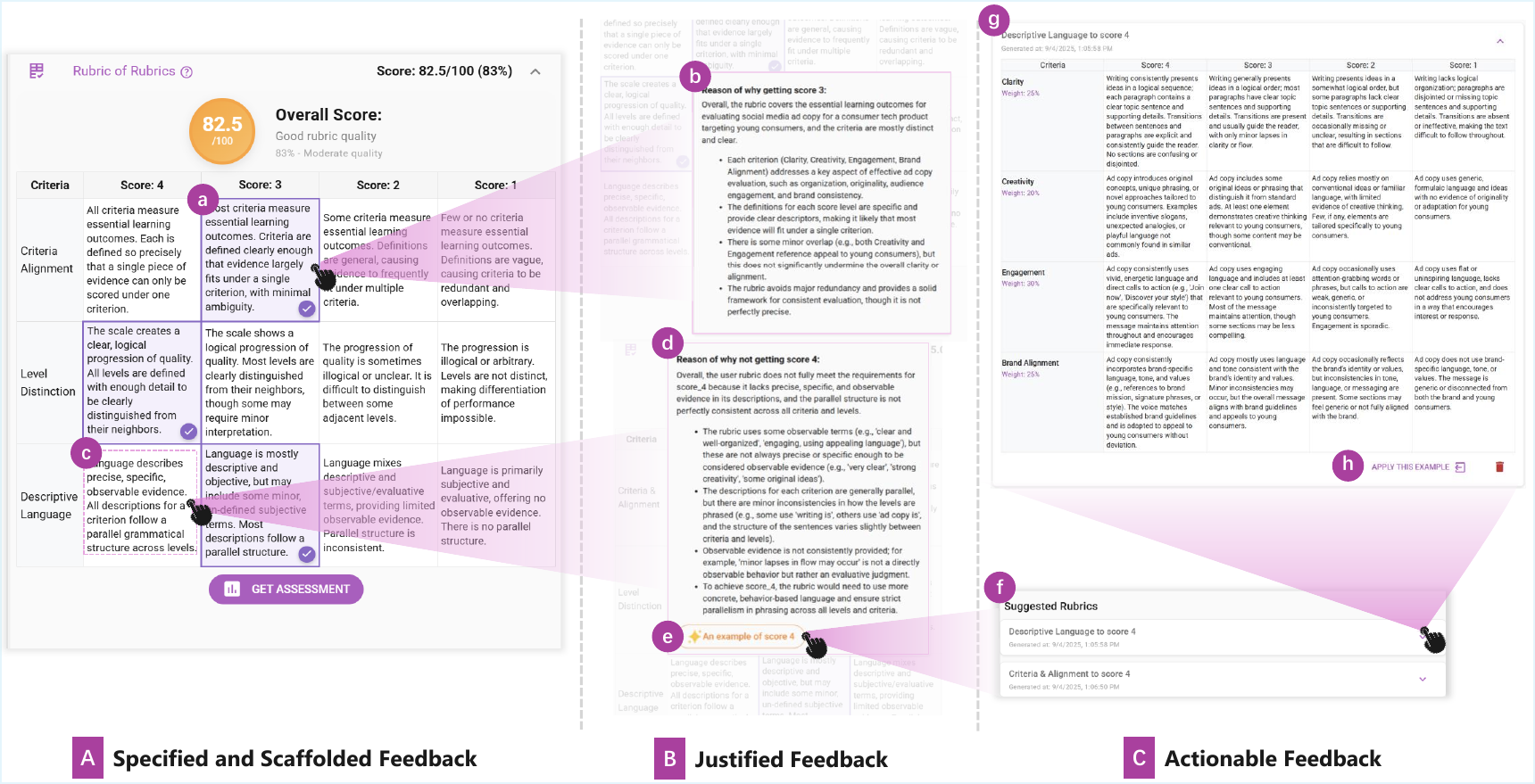}
    \caption[]{The interactive workflow in iRULER Rubric Creation User Interface.
    (a) An initial assessment on the rubric-of-rubric provides \textit{Specific} and \textit{Scaffolded} feedback on the user's rubric. Users then get \textit{Justified} feedback via (b) \textit{Why} and (d) \textit{Why Not} explanations. Requesting (e) a \textit{How To} example generates (f$-$g) a full suggested rubric for side-by-side comparison, which can be adopted with (h) \textit{Apply This Example} button.
    }
    \Description{A three-phase diagram illustrates the iRULER rubric creation workflow. 
    Phase A (left, labeled "Specified and Scaffolded Feedback") shows the "Rubric of Rubrics" meta-rubric table with an overall score of 82.5/100. For both the "Criteria Alignment" and "Descriptive Language" criteria, score 3 is highlighted in purple with a checkmark. A cursor points at the cell of (Criteria Alignment, Score 3), expanding into Phase B's "Reason of why getting score 3" pop-up box. Another cursor points at (Descriptive Language, Score 4), expanding into Phase B's "Reason of why not getting score 4" pop-up box.
    Phase B (middle, labeled "Justified Feedback") displays two pop-up boxes with detailed feedback. The top box, titled "Reason of why getting score 3" contains bullet points explaining the score. The bottom box, titled "Reason of why not getting score 4," explains the deficiencies and includes a button labeled "An example of score 4." A cursor points at this button, which expands into Phase C's "Suggested Rubrics" list. 
    Phase C (right, labeled "Actionable Feedback") displays a "Suggested Rubrics" list, containing sample rubric categories like "Descriptive Language to score 4" and "Criteria Alignment to score 4". A cursor points at a drop-down indicator for the first item in the list, expanding into a detailed rubric sample for "Descriptive Language to score 4".}
    \label{fig:rubric-flow}
\end{figure*}

Starting with a blank table, the user renames the rubric (Fig.~\ref{fig:rubric-interface} A1) and specifies the overall Task Description (Fig.~\ref{fig:rubric-interface} A2) to give context to the rubric.
Four levels (1--4 points) are offered by default; users can adjust the scale between a minimum of 3 and a maximum of 6 levels (Fig.~\ref{fig:rubric-interface} A3, \textbf{DG2: Scaffolded}).
To begin populating the rubric, the user can enter a criterion (Fig.~\ref{fig:rubric-interface} A4) to specify the basis on which to precisely judge writing (\textbf{DG1: Specific}).
Upon naming the criterion, the user can specify the details of each level by describing them in text.
Since writing descriptions is tedious and users may be unable to articulate details or distinguish between levels, we offer AI suggestions (See in Fig. \ref{fig:ai-assist-features}) based on LLM prompting (see Appendix~\ref{appendix:rubric-design-prompts} for details).
After adding multiple criteria, the user can specify the weight (\textbf{DG2: Scaffolded}) for each criteria toward an overall score.

\subsubsection{Rubric Feedback Panel}
The components of Rubric Feedback are similar to Writing Revision, using the same 
overall score calculation, rubric table, and \textit{Why}/\textit{Why Not} justifications..

One difference is that counterfactual alternative (\textit{How To}) rubrics fully replace the rubric in the Rubric Design Panel. We do not track changes to avoid overwhelming the user with a complex UI.
Since this feedback helps to improve the quality of the rubric for writing revision, it qualifies the latter rubric (\textbf{DG5: Qualified}).

\subsubsection{Walkthrough}
We illustrate the rubric creation workflow through Clara, a user designing her own evaluation criteria (Fig. \ref{fig:rubric-flow}). This process ensures the user-defined criteria itself is qualified \textbf{(DG5: Qualified)}.
After Clara's drafted rubric receives an initial assessment against a meta-rubric, she got a \textbf{Specific} and \textbf{Scaffolded} feedback. She begins to explore the \textbf{Justified}  feedback, clicking cells to see (b) \textit{Why} her rubric received a certain score and (d) \textit{Why Not} it failed to meet a higher standard.
Building on this diagnosis, she requests (e) a \textit{How To} counterfactual example for actionable guidance (\textbf{DG4: Actionable}, see Appendix~\ref{appendix:how-to-rubric-prompt} for prompt details). Unlike the writing application, the (f) suggested rubric is displayed holistically in (g) a dedicated area for Clara to review and adopt it with (h) a single click.

Next,
Clara applies her rubric in the writing revision UI (Fig.~\ref{fig:writing-interface}). 
As she writes with the rubric feedback, 
she may disagree with certain scores, see many borderline drafts clustered at the same level, or realize that important qualities (e.g., Creativity or Engagement) have not been accounted for. 
In such cases, she can click the top-right \textit{Edit Rubric} button to return to the rubric creation UI, refine the wording, thresholds, or weights of her criteria, and re-apply the updated rubric.
She can then perform these steps altogether in an iteratively refinable manner \textbf{(DG6: Refinable)}.



\subsection{iRULER Implementation}
\name{} was implemented as a web application using a serverless architecture. The front-end, built with Vue.js\footnote{https://vuejs.org/} and the Quasar component library\footnote{https://quasar.dev/}, manages all application logic and directly interacts with the OpenAI GPT-4.1 API. 
The Writing Revision application uses quasar's WYSIWYG editor enhanced with a custom change-tracking system built upon the jsdiff library\footnote{https://github.com/kpdecker/jsdiff}. 
The Rubric Creation application features a custom, table-based user interface with embedded AI-assistive features. 
The Justified (\textit{Why, Why Not}) and Actionable (\textit{How To}) feedback were generated from LLM with structured prompt templates (see Appendix~\ref{appendix:prompts}).
For details of the rubric JSON format used in our system, see Appendix~\ref{appendix:rubric-json}.

To formalize the recursive nature of \name{}, we model its evaluation process as a two-stage conditional generation problem.
In both stages, the same \textsc{LLM} is invoked with a \textbf{unified prompt template} (Appendix~\ref{appendix:iruler-evaluation-prompts}), so that a common evaluative logic can be applied to different artifact types.

In the \textcolor{irulerPurple}{Rubric Qualification Loop}, the model refines the \textcolor{irulerPurple}{user's draft rubric} by evaluating it against the rubric-of-rubrics and the rubric revision goal, producing a \textcolor{irulerPurple}{qualified, revised rubric}.

\begin{equation*}
    \textcolor{irulerPurple}{\text{Revised Rubric}} = \text{LLM}\left(
    \begin{matrix}
        \textcolor{irulerPurple}{\text{Draft Rubric}}, \\
        \text{Rubric-of-Rubrics}, \\
        \text{Revision Goal for } \textcolor{irulerPurple}{\text{Rubric}}
    \end{matrix}
    \right)
\end{equation*}


In the \textcolor{irulerBlue}{Writing Revision Loop}, the model uses this \textcolor{irulerPurple}{revised rubric} along with the writing revision goal to evaluate the \textcolor{irulerBlue}{user's draft} \textcolor{irulerBlue}{writing}, producing a piece of \textcolor{irulerBlue}{revised writing}.
\begin{equation*}
    \textcolor{irulerBlue}{\text{Revised Writing}} = \text{LLM}\left(
    \begin{matrix}
        \textcolor{irulerBlue}{\text{Draft Writing}}, \\
        \textcolor{irulerPurple}{\text{Revised Rubric}}, \\
        \text{Revision Goal for } \textcolor{irulerBlue}{\text{Writing}}
    \end{matrix}
    \right)
\end{equation*}




\subsection{Technical Validation of iRULER Scoring Metric}

To evaluate the consistency and human-faithfulness of \name{}'s automatic scoring, we conducted two analyses of iRULER's predicted rating on 
i) the variability of the stochastic scoring, and
ii) how similarly human expert ratings match iRULER.
We performed these analyses on the ICNALE Edited Essays dataset~\cite{ishikawa2018icnale}, 
which has 640 essays of 200--300 words from ESL students, scored by human raters.
For each essay, we run \texttt{gpt-4.1} with our prompt (temperature $=0$) five times\footnote{Fixing a low temperature reduces randomness, while five runs provide a practical trade-off between capturing residual stochasticity and token cost, a practice aligned with recent work on evaluating LLM consistency~\cite{zhou2025explainable, szymanski2025limitations, cheng2024relic}.}, 
obtaining five sets of criteria scores and a 0--100 total score using the same rubric as in our user study (see Table~\ref{tab:esl-profile-rubric} in Appendix~\ref{appendix-user-study-rubric}).

\subsubsection{Evaluating Intra-Model Reliability}
\label{sub-sec:llm-internal-consistency}


We treat the five runs per essay as five raters applying the same rubric and compute Krippendorff’s $\alpha$\footnote{Krippendorff’s $\alpha$ is widely used to quantify inter-rater reliability, including in assessments involving LLMs and humans~\cite{bojic2025comparing, pang2025understanding, ma2025should}. Here, we apply ordinal for criteria scores, interval for the total score.}
and mean with-in essay standard deviations.
As shown in Table~\ref{tab:tech_llm_judge}, 
the model shows minimal stochastic fluctuation\footnote{$\alpha$ values above .80 are considered reliable for drawing conclusions~\cite{krippendorff2018content,krippendorff2011computing}.}, 
all criteria achieve $\alpha \ge .82$, and the total score reaches $\alpha = .95$, 
with SDs of 0.06--0.11 on four criteria and 1.7 on the 0--100 total score.

\begin{table}[t]
\centering
\small
\setlength{\tabcolsep}{4pt}
\caption[]{Results of technical validation of the \name{} scoring metric. 
}
\label{tab:tech_llm_judge}
\begin{tabular}{lcccccc}
\toprule
& \multicolumn{3}{c}{Intra-Model Reliability} 
& \multicolumn{2}{c}{Expert-LLM Agreement} \\
\cmidrule(r){2-4} \cmidrule(l){5-6}
Dimension 
& $\alpha$ & 95\% CI & SD 
& QWK & 95\% CI \\
\midrule
Content        & 0.82 & 0.79--0.85 & 0.11 & 0.52 & 0.47--0.57 \\
Organization   & 0.86 & 0.83--0.89 & 0.07 & 0.57 & 0.52--0.61 \\
Vocabulary     & 0.90 & 0.88--0.92 & 0.07 & 0.61 & 0.57--0.66 \\
Language Use   & 0.92 & 0.90--0.94 & 0.06 & 0.66 & 0.62--0.70 \\
\midrule
Total score    & 0.95 & 0.95--0.96 & 1.69 & 0.68 & 0.64--0.72 \\
\bottomrule
\end{tabular}
\end{table}



%
\subsubsection{Evaluating Expert-LLM Score Agreement}
\label{sec:dataset-validation-human-ai}

We average the five runs per essay and compute Quadratic Weighted Kappa (QWK)\footnote{QWK was the official evaluation metric in the Hewlett Foundation’s ASAP automated essay scoring competition: \url{https://www.kaggle.com/c/asap-aes}.} between these scores and ICNALE expert scores.
As shown in Table~\ref{tab:tech_llm_judge}, 
Expert-LLM agreement is \emph{moderate to substantial}\footnote{$\kappa$ values in .41--.60 indicate moderate agreement; .61--.80 substantial; and above .80 almost perfect agreement~\cite{landis1977measurement}.}:
$QWK = .68$ on the 0--100 total score, with per-criterion QWK ranging from .52 (Content) to .66 (Language Use),
comparable to recent LLM-based AES systems and typical human–human agreement on fine-grained writing scores~\cite{tang2024harnessing,jonsson2007use,uchida2024evaluating}.

Following recent HCI work on LLM-as-a-judge~\cite{shankar2024validates, szymanski2025limitations}, we do \emph{not} intend for \name{} to replace human experts.
Instead, we use it as a consistent, rubric-aligned proxy that achieves substantial agreement on overall scores and moderate-to-substantial agreement on criteria, enabling repeatable evaluation in our user studies.

\section{Evaluation}

To evaluate \name{}, we conducted: 
(i) two controlled experiments assessing \textbf{Writing Revision} and \textbf{Rubric Creation} separately\footnote{For scalability and affordability, we assessed effectiveness using LLM-as-a-judge, instead of human raters that would be require numerous human ratings and be excessively costly.}; 
(ii) a qualitative study examining the \textbf{End-to-End} feedback loop; and 
(iii) a validation of our LLM-as-a-judge metric against expert human raters, ensuring the reliability of our evaluation method. All studies were approved by our Institutional Review Board (IRB).



\textbf{Controlled Study Design.} To precisely measure the \name{}'s impact and avoid confounds, we decoupled the writing and rubric tasks into two separate between-subjects user studies. Participants either revised an essay or designed a rubric from scratch, and were randomly assigned to one of three feedback conditions (Table~\ref{tab:conditions}).



\begin{itemize}
    \item \textbf{Text-LLM.} Participants saw only an Overall Score (0--100) and a generic LLM-generated comment. Without any specific criteria and or rubric scaffolding; see Fig.~\ref{fig:writing-baseline-text},~\ref{fig:rubric-baseline-text}. To get \textit{actionable} revision suggestions, participants had to manually prompt the LLM via a text chatbox.

    \item \textbf{Rubric-LLM.} Participants saw the rubric with \textit{specific} criteria and \textit{scaffolded} performance levels, and per-criterion scores. 
    Each criteria is rated as a Subscore (e.g. 1--4), and weighted (e.g. 30\%, 20\%) based on priority to compute the Overall Score (0--100).
    This rubric is read-only: it did not \textit{justify} the scores.
    Participants get \textit{actionable} revision suggestions with LLM chat like Text-LLM; 
    see Fig.~\ref{fig:writing-baseline-rubric},~\ref{fig:rubric-baseline-rubric}.

    \item \textbf{\name{}.} Participants saw the same rubric and weighted Overall Score as in Rubric-LLM, with additional interactivity in the rubric table.
    Clicking on scored or alternative levels revealed \textit{justifications} for Why a rating was given or Why Not another rating.
    Clicking on alternative levels' example buttons invokes a criteria-specific \textit{actionable} revision to achieve the target level; see Fig.~\ref{fig:writing-flow},~\ref{fig:rubric-flow}.

\end{itemize}
In the \textbf{End-to-End} study, participants only used the iRULER system and no baseline.

\begin{table}[t]
  \centering
  \small
  \setlength{\tabcolsep}{4pt}
  \caption[]{
    Comparison of \name{} against baselines (Text-LLM, Rubric-LLM) on DG1--4 
  (Specific, Scaffolded, Justified and Actionable).
  DG5--6 (Qualified and Refinable) pertain to the end-to-end rubric creation and feedback workflow.
    }
  \label{tab:conditions}
  \begin{tabular}{lccc}
    \toprule
    DGs    & Text-LLM & Rubric-LLM   & \name{} (ours) \\
    \midrule
    DG1     & \textcolor{lightgray}{Generic}           & Per Criteria     & Per Criteria \\
    DG2   & \textcolor{lightgray}{Open-ended}            & Rubric            & Rubric \\
    DG3    & \textcolor{lightgray}{Generic Text} & \textcolor{lightgray}{N.A.}            & Intelligible Why \& Why Not \\
    DG4   & \textcolor{lightgray}{via Chat prompt} & \textcolor{lightgray}{via Chat prompt} & Interactive Counterfactual \\
    \bottomrule
  \end{tabular}
\end{table}

\subsection{Research Questions and Hypotheses}

Our evaluation was designed to answer these research questions:
\begin{enumerate}
    \item[{RQ1.}] \textbf{Performance:} How does \name{}'s feedback impact the quality of the final artifacts in writing revision and rubric design tasks?
    \item[{RQ2.}] \textbf{Efficiency: } How does \name{} affect the effort and speed to produce the final artifacts?
    \item[{RQ3.}] \textbf{Perceived Qualities:} How do users interpret and gauge the feedback's helpfulness, correctness, and their own sense of control in \name{}?
    \item[{RQ4.}] \textbf{Confidence:} How does interacting with \name{} affect users' perceived confidence?
    \item[{RQ5.}] \textbf{Skill Transfer:} 
    Does using \name{} in the writing revision task lead to transferable improvements in users' unaided revision skills, even though it is not primarily designed as a tutoring tool?
\end{enumerate}
%
Specifically, we hypothesize:
\begin{enumerate}
\item[{H1.}]{\textbf{Performance.}}
Compared to baselines, the final artifacts produced by participants with \name{} should score higher and better satisfy predetermined criteria. 

\item[{H2.}]{\textbf{Efficiency.}}
Compared to baselines, participants using \name{} will be fastest to iterate toward their final artifact and should make the fewest iterations.

\item[{H3.}]{\textbf{Perceived Qualities.}}
Compared to baselines, participants would rate \name{} as most \textit{helpful},  most \textit{correct} (aligned with their goals), and most \textit{controllable}.

\item[{H4.}]{\textbf{Confidence.}}
Compared to baselines, participants would have the greatest improvement in confidence levels after using \name{}.

\item[{H5.}]{\textbf{Skill Transfer.}}
In the Writing Revision experiment, participants using \name{} will show larger improvement in unaided revision skills than those in the baseline conditions.

\end{enumerate}

\subsection{Evaluating iRULER for Writing Revision}
\label{sub-sec:writing-revision-eval}
This experiment focused on evaluating how well \name{} helps users to improve the quality of written essays.

\subsubsection{Materials}
We selected four argumentative essays from the ICNALE Edited Essays dataset~\cite{ishikawa2018icnale}. 
This dataset contains essays scored by expert raters using the ESL Composition Profile~\cite{jacobs1981testing}, a well-established public rubric,
which provides criterion-level scores.
We used four\footnote{We omitted the Mechanics criterion to simplify the number of criterial participants need to consider, since it which was only weighted 5\% in the original ESL Composition Profile.} criteria: Content, Organization, Vocabulary, and Language Use (details in Appendix~\ref{appendix-user-study-rubric} Table~\ref{tab:esl-profile-rubric}).
To ensure a consistent starting point with enough room for improvement, all selected essays are at a medium level (approx. 60/100 score). 
The essays covered two topics, which we refer to as the \textbf{Part-Time Job}\footnote{Essay topic: ``It is important for college students to have a part-time job.''} and the \textbf{Smoking Ban}\footnote{Essay topic: ``Smoking should be completely banned at all the restaurants in the country.''} topics.
Essays were 197--205 words long.

\subsubsection{Procedure}

We conducted a 60-minute remote study via Zoom. After providing consent and completing a background survey, participants were randomly assigned to one of three conditions ($N=16$ each). They received a tutorial and a screening quiz before the main session. We employed a think-aloud protocol, recording screen and audio for analysis.

The main session consisted of three stages, with the assignment of the Part-Time Job and Smoking Ban topics counterbalanced across participants:
\begin{enumerate}
\item \textbf{Pre-Task:} Participants revised one essay without AI assistance to establish their baseline revision skill.
\item \textbf{Main Task:} Participants used their assigned Feedback Type to revise two new essays on the alternate topic, iterating until satisfied. After each revision, they completed a short survey, rating the feedback's perceived \emph{helpfulness}, \emph{correctness}, and their \emph{sense of control}.
\item \textbf{Post-Task:} Participants revised a final essay on the initial topic, again without AI assistance, to measure skill transfer.
\end{enumerate}
The session concluded with a post-task questionnaire on feature helpfulness and confidence, followed by a brief semi-structured interview to gather qualitative insights. 
Participants were compensated \$15.60 USD equivalent in local currency.
All data were anonymized and stored securely in compliance with IRB protocols.


\subsubsection{Measures}
To test our hypotheses, we measured:
\begin{enumerate}
    \item \textbf{Quality Scores} based on LLM-as-a-Judge metrics using Text-based LLM, and our Rubric-based LLM (see Appendix~\ref{appendix:prompts} for prompt details). 
    We used the two metrics to check if our Rubric-based LLM scoring is consistent with the common Text-based LLM scoring approach.
    We also measured \textbf{Criteria Subscore Levels} from the Rubric LLM.
    We calculated the score improvement $\Delta$Score from initial essay to final revised essay for the pre-, main, and post-tasks.
    Specifically for this experiment, we calculated skill transfer as the difference between a participant's performance gain in the post-task and their baseline performance gain in the pre-task (both conducted without AI assistance).

    \item \textbf{Usage Logs} of \textit{task time}, \textit{number of iterations}, and \textit{usage} of Intelligibility features \textit{(Why, Why Not, How-to)}.

    \item \textbf{Perceived Ratings} on \textit{Helpfulness}, \textit{Correctness}, and \textit{Sense of Control} regarding each Feedback Type, and 
    self \textit{Confidence} in \textit{Writing}, \textit{Reviewing}, \textit{Revising} and \textit{Using Rubrics}.
    All ratings were measured along a 7-point Likert scale ($-3$ Strongly Disagree to $+3$ Strongly Agree).
\end{enumerate}
%

\subsection{Evaluating iRULER for Rubric Creation}
\label{sub-sec:rubric-creation-eval}
This experiment focused on evaluating how well \name{} helps users to create high-quality rubrics.
Unlike the Writing Revision Experiment, the task is to create rubrics from scratch, not to improve existing ones.
So it omits a pre/post-test on skill transfer.

\subsubsection{Materials}
\label{sub-sec:rubric-creation-materials}
We selected two distinct topics on which participants designed rubrics, drawing on established practices in evaluating social media ad copy~\cite{smith2017assessing,belch2016advertising} and student public speaking performances~\cite{stevens2023introduction}.
For the rubric feedback, we used the rubric-of-rubrics~\cite{arter2000rubrics,stevens2023introduction}
that defines a rubric to evaluate rubrics. 
It
comprises three criteria---Criteria Alignment, Level Distinction, and Descriptive Language---each rated on a 4-point scale (see Table~\ref{tab:rubric-of-rubrics} in Appendix~\ref{appendix-user-study-rubric} for criteria and level descriptions).


\subsubsection{Procedure}
This experiment followed the same initial protocol as the Writing Revision experiment. 
Participants ($N=12$ per condition) were randomly assigned to one of three conditions to design two rubrics from scratch, with the genre order randomized, 
and iterating with the system's feedback until satisfied. 
Upon completing each task, participants immediately rated the feedback's \textit{helpfulness}, \textit{correctness}, and their \textit{sense of control}. The session concluded following the same protocol as the previous experiment.



\subsubsection{Measures}
We measured the same metrics as in the Writing Revision Experiment to allow for direct comparison. The only adaptation was 
modifying self \textit{Confidence} skills to be relevant to this task, omitting \textit{Writing} and including \textit{Designing Rubrics} instead.

\subsection{Evaluating iRULER End-to-End (Qualitative)}
\label{sub-sec:end-to-end-eval}

This experiment qualitatively examines how users can create, use, and refine rubrics for writing revision.
This specifically investigates the benefit and challenges toward constructing qualified rubrics (\textbf{DG5:~Qualified}) and refining them (\textbf{DG6:~Refinable}).


\subsubsection{Materials} 
We employed two tasks in this experiment to balance control with coverage of diverse genres. 
The first was a \textit{fixed-topic} ``Social Media Ad Copy'' task with a fixed sample text\footnote{The sample ad copy was adapted from posts in the r/copywriting subreddit: \url{https://www.reddit.com/r/copywriting/}.}, 
chosen from our Rubric Creation experiment (Section~\ref{sub-sec:rubric-creation-materials}) as it yielded positive results: 
its short, constrained format let participants quickly learn the system and complete multiple revision cycles within a single session.
The second was an \textit{open-ended} task---allowing participants to select any genre of writing they wished to evaluate---to probe how well \name{}'s rubric-based workflow generalizes beyond our predefined experimental topics.

\subsubsection{Procedure}

For both tasks, participants first drafted a rubric from scratch using the Rubric Creation UI (Section~\ref{sub-sec:rubric_creation_system}), then applied it to a piece of writing in the Writing Revision UI (Section~\ref{sub-sec:writing_feedback_system}).
They were then free to refine their rubrics by returning to the creation interface to update criteria and re-evaluate the text. 
Participants were asked to think aloud during the tasks.
Each session lasted approximately 45 minutes.
We gratefully acknowledge the participants who volunteered their time for this particular experiment.

\subsubsection{Measures}
We only investigated the iRULER feedback condition, since it was explicitly designed to support user-defined rubrics.
We measured \textit{Quality Scores} and \textit{number of revision cycles} in this experiment.
Quality Scores were measured as per the previous two experiments, while the number of revision cycles was determined by excluding the initial round of rubric creation and application\footnote{E.g. 2 cycles $=$ initial design \& apply + \textit{(redesign \& reapply)} + \textit{(redesign \& reapply)}}.

\subsection{Validating LLM-as-a-Judge Metric}
\label{sec:human-expert-validation}


Since we had used LLM-as-a-judge (Text-LLM and Rubric-LLM Scores) to scalability and affordability score the writing artifacts from various conditions, we analyze the validity of these metrics by comparing against two human expert raters.
This examines if the LLM scores merely reflect text-model alignment, rather than actually reflecting human-perceived work quality.
Due to the difficulty of recruiting rubric experts, we focused on the Writing Revision task with expert English tutors.

\subsubsection{Materials}
We randomly sampled 96 essays (50\% of the full set of 192), with a balanced distribution across the two topics and three feedback conditions 
produced in the Writing Revision experiment. Essays were 194--310 words long.
To estimate inter-rater reliability, the same 24 essays were assigned to both experts, while the remaining 72 essays were divided evenly. 
Each expert rated 60 essays in total (24 shared and 36 unique).

\subsubsection{Procedure}
Before the experiment, we held a brief online orientation session to explain the protocol; 
raters then independently completed all scoring asynchronously within one day at their own pace using the same rubric as in our Writing Revision Experiment (Section~\ref{sub-sec:writing-revision-eval}).
Participants were not informed of the various feedback conditions, thus would not know if each essay were revised based on different feedback types. 
They were compensated \$70.50 USD (equivalent in local currency) upon completion.

\subsubsection{Measures}
Same as Section~\ref{sec:dataset-validation-human-ai}, we used Krippendorff’s $\alpha$~\cite{krippendorff2011computing} to estimate inter-rater reliability between the two experts. 
We then assessed Expert-LLM alignment by averaging the two experts’ scores to obtain a ground-truth score per essay and computing the QWK between the Text-LLM and Rubric-LLM scores and this ground truth for each criterion and the total score.

\section{Results}
\label{sec:results}
We present the results of our evaluation, organized by our hypotheses. 
We fit a linear mixed-effects model to each dependent variable (measure) with Feedback Type and \# Iterations as fixed main effects, and Participant and Task ID as random effects to account for individual variance.
For specific comparisons, we performed post-hoc contrast t-tests and report significant results.
We apply a significance level of $\alpha = .002$ to account for up to 25 multiple comparisons up and avoid Type I error\footnote{Type I error $=$ false positive (rejecting a true null). To control family-wise error with multiple comparisons, we applied a Bonferroni correction (\(\alpha'=\alpha/m\)); here, \(0.05/25=0.002\).}.
We also report thematic findings from post-experiment interviews to contextualize our quantitative results while incorporating the qualitative findings from our end-to-end evaluations.
Finally we report the results of the Expert-LLM alignment analysis that validates our LLM-as-a-judge metrics.



\subsection{Writing Revision Experiment}

We recruited 48 participants (38 female, 10 male; 19--28 years old, \textit{M} $=$ 22.3, \textit{SD} $=$ 2.0) from our university's online recruitment platform. All were native or full-professional proficiency English speakers, comprising 28 undergraduate and 20 postgraduate students. Participants reported a range of writing frequencies, with most writing at least a few times a month (31/48) or weekly (9/48). 
Their familiarity with rubrics varied: 10.4\% not familiar at all, 20.8\% somewhat familiar, 58.3\% familiar, 10.4\% very familiar.

We found that \name{} significantly improved writing quality (H1), efficiency (H2), perceived qualities (H3), writing confidence (H4), and skill transfer (H5) compared to Text-based LLM feedback.
Non-intelligible Rubric-based LLM feedback was in-between them.

\begin{figure}[t]
  \centering
  \includegraphics[width = 5.1cm]{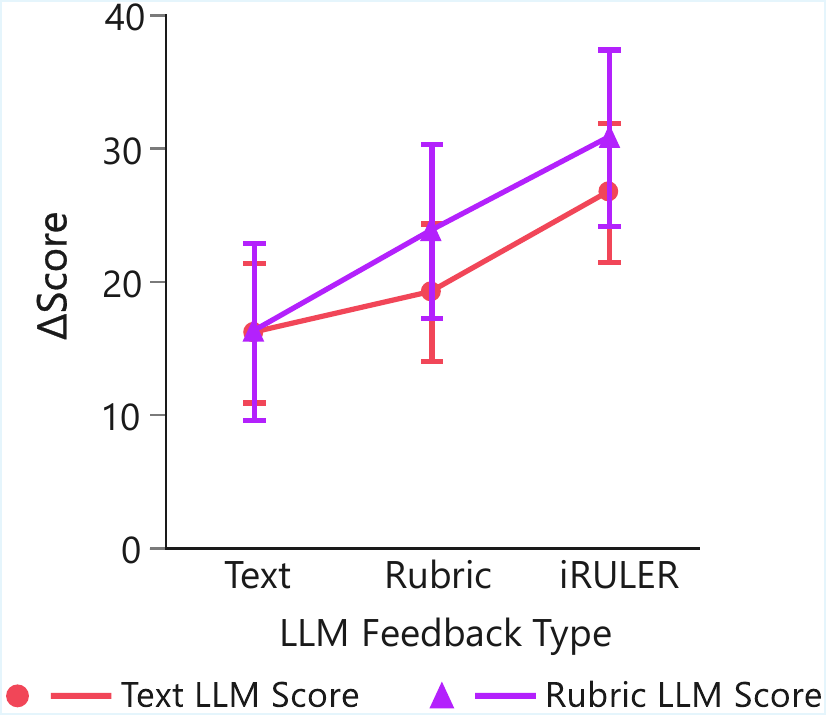}
  \caption[]{Result of the writing quality improvement ($\Delta$Score) as assessed by text-based and rubric-based LLM evaluators. 
  A significant interaction indicates iRULER's superiority in the Main task.
  Error bars represent 95\% confidence intervals, here and for all result figures.}
  \Description{A line graph showing the improvement in writing quality score on the Y-axis against the three LLM Feedback Types (Text, Rubric, iRULER) on the X-axis. The Y-axis, labeled "Delta Score," ranges from 0 to 40. Two line plots, both with an increasing trend from Text to iRULER, are shown. 
  The purple line with triangles, representing the "Rubric LLM Score," shows mean values of 16.2 for the Text condition, 23.8 for the Rubric condition, and 30.8 for the iRULER condition. 
  The red line with circles, for the "Text LLM Score," shows a similar increasing trend with mean values of 16.1 for Text, 19.2 for Rubric, and 26.7 for iRULER. For both line plots, the iRULER condition shows the highest mean score improvement. Error bars are present for all data points.}
  \label{fig:writing-score}
\end{figure}

\subsubsection{Writing Quality (H1)} 
For the following analysis we use our LLM scoring method to score each essay five times and calculate the mean values. 
Our analysis later in Section~\ref{sub-sec:llm-internal-consistency} verified the human-alignment of this metric.

\textit{$\Delta$Score.} 
Fig. \ref{fig:writing-score} shows the Score improvements measured by Text-based LLM and Rubric-based LLM for the main tasks.
We found a significant main effect of Feedback Type (\textit{p} $<$ .0001) on both types of score improvements. 
\name{}'s score improvements (\textit{M}\textsubscript{text} $=$ 26.7, \textit{M}\textsubscript{rubric} $=$ 30.8) were significantly higher than in Text-LLM (\textit{M}\textsubscript{text} $=$ 16.1, \textit{M}\textsubscript{rubric} $=$ 16.2, both \textit{p} $<$ .0001).
\name{} was only significantly higher than in Rubric-LLM on the text-based score improvement (\textit{M}\textsubscript{text} $=$ 19.2, \textit{p} $=$ .0003), not on the rubric-based score improvement (\textit{M}\textsubscript{rubric} $=$ 23.8, \textit{p} $=$ n.s.).

\begin{figure*}[t]
  \centering
  \includegraphics[width = 16.5cm]{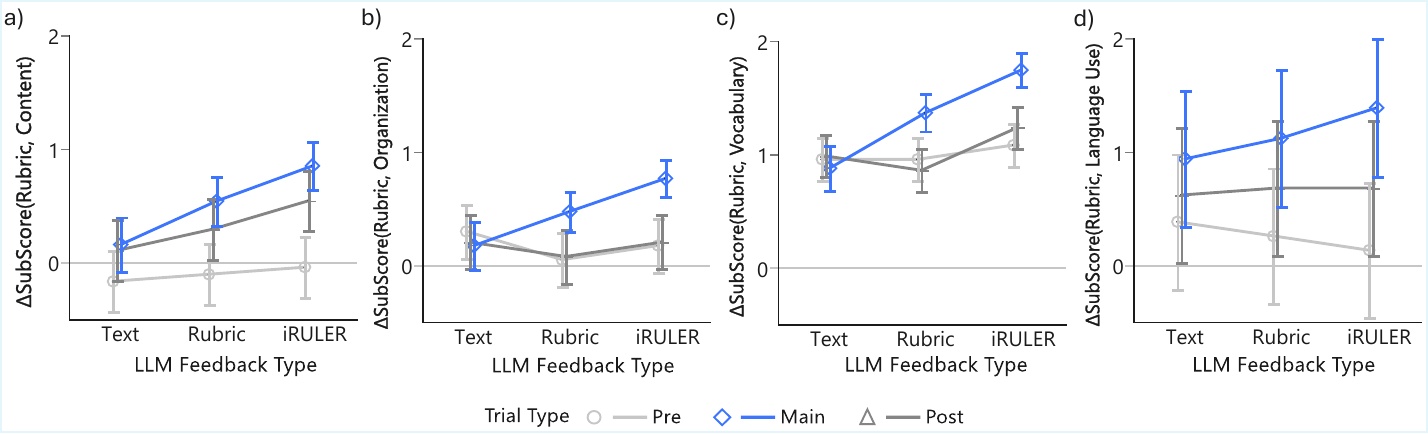}
  \caption[]{Criterion-level improvements ($\Delta$SubScoreLevel) for the Writing Revision experiment. The performance gains across all criteria: a) Content, b) Organization, c) Vocabulary, and d) Language Use, showing a consistent pattern, where the \name{} condition led to the greatest performance gains in the Main task.}
  \Description{Four separate line graphs (a-d) arranged 1x4 showing the ΔSubScoreLevel on the Y-axis against three LLM feedback types (left to right: Text, Rubric, iRULER) on the X-axis. Each graph represents a different criterion: a) Content, b) Organization, c) Vocabulary, and d) Language Use. The y-axis for all plots ranges from 0 to 2 in increments of 1. 
  Each line graph contains three line plots representing Trial Types, each shown with different colors and markers: Pre (gray circles), Main (blue diamonds), and Post (gray triangles). Error bars at each point indicate 95\% confidence intervals. 
  Each line graph observes a similar pattern: the Main trial types have lines that are consistently above the lines of Post and Pre trial types, and show an increasing trend from Text to iRULER; Post trial types have lines above or similar to the lines of Pre trial types, both of which are generally flat lines.
  The scales on each line graph are different, with a) Content ranging from below 0 to 1, b) Organization ranging from 0 to 1, c) Vocabulary ranging from below 1 to 2, and d) Language Use ranging from below 0 to 2. 
  Error bars are present for all data points.}
  \label{fig:writing-subscore}
\end{figure*}

\textit{$\Delta$Subscore.}
Next, we examine which criteria was particularly helped by \name{} by analyzing the criteria subscore levels.
Fig. \ref{fig:writing-subscore} shows the Criteria Subcore improvements measured by the Rubric-based LLM for the pre-, main, and post-tasks.
We found a significant interaction between Feedback Type and Trial Type for the Organization and Vocabulary criteria (all \textit{p} $<$ .001), but not for Content nor Language Use. 
For main tasks, 
\name{}'s score improvement was significantly higher than in Text-LLM for Content, Organization, Vocabulary and Language Use criteria (all \textit{p} $\le$ .001), 
and higher than in Rubric-LLM for Content and Vocabulary (all \textit{p} $\le$ .002).


\subsubsection{Efficiency (H2)}
\label{sub-sub-section-efficiency}
\begin{figure}[t]
  \centering
  \includegraphics[width = 4.2cm]{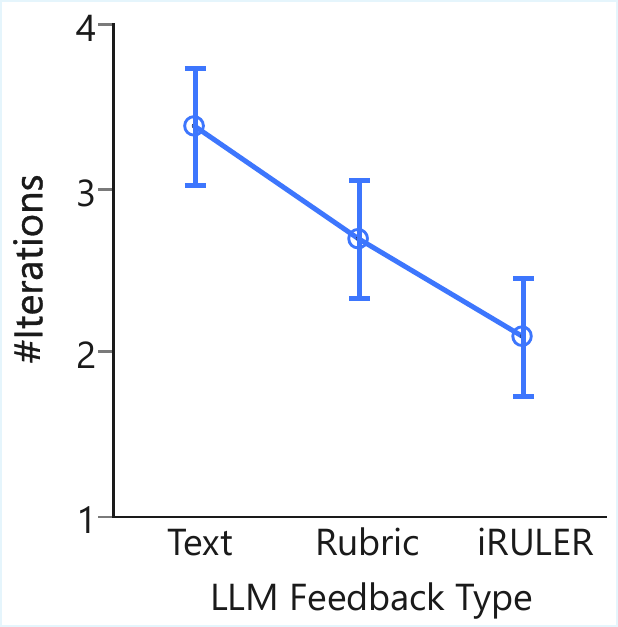}
  \caption[]{Average number of iterations in the Main task of Writing Revision experiment.
  \name{} participants achieved better results with significantly greater efficiency.
  }
  \Description{A line graph showing the average number of iterations on the Y-axis against the three LLM Feedback Types (left to right: Text, Rubric, iRULER) on the X-axis. The Y-axis, labeled "Number of Iterations," ranges from 1 to 4. A single line plot in blue, with circular markers, shows a clear downward trend from Text to iRULER. The mean value is approximately 3.4 for the Text condition,  approximately 2.7 for the Rubric condition, and approximately 2.1 for the iRULER condition. Error bars are present for all data points.}
  \label{fig:writing-iteration}
\end{figure}

Beyond improving quality, \name{} also made the writing process more efficient with fewer iterations, although there was no difference in task speed between conditions.
Fig. \ref{fig:writing-iteration} shows the \# Iterations across Feedback Types.
There was a significant main effect of Feedback Type on \# Iterations ($p < .0001$), with \name{} requiring the fewest iterations (\textit{M} $=$ 2.09) over Text-LLM (\textit{M} $=$ 3.38, \textit{p} $<$ .0001) but not Rubric-LLM (\textit{M} $=$ 2.69, \textit{p} $=$ n.s). 

\subsubsection{Perceived Qualities (H3)}
\begin{figure}[t]
  \centering
  \includegraphics[width = 5.5cm]{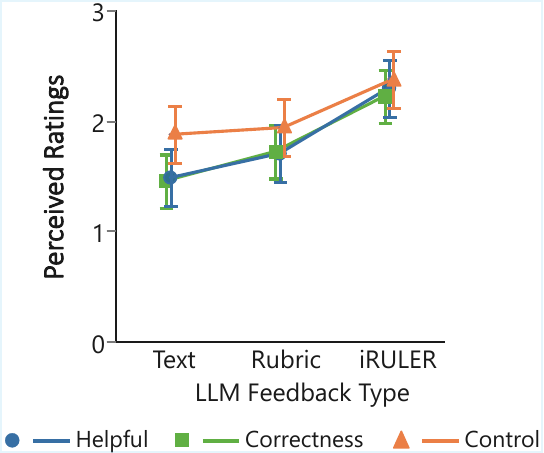}
  \caption[]{
  Per-trial subjective ratings for the Writing Revision experiment, showing \textit{least-squares means} for Helpfulness, Correctness, and Control. \name{} showed significantly superior performance compared to baselines.
  }
  \Description{
  A line graph plotting "Perceived Ratings" (Y-axis, range 0 to 3) against three "LLM Feedback Types" (X-axis: Text, Rubric, iRULER). Three color-coded lines represent the metrics, all showing an upward trend with error bars:
  1. Helpful (blue line, circle markers): Starts around 1.5 for Text, rises to 1.75 for Rubric, and peaks at roughly 2.3 for iRULER.
  2. Correctness (green line, square markers): Follows a similar path to Helpful, starting around 1.5 for Text, rising to 1.75 for Rubric, and peaking at roughly 2.2 for iRULER.
  3. Control (orange line, triangle markers): Starts higher than the others at approx 1.9 for Text, rises slightly to 2.0 for Rubric, and finishes highest at approx 2.4 for iRULER.
  Overall, the iRULER condition achieves the highest ratings across all three subjective measures.
  }
  \label{fig:writing-perceived-ratings}
\end{figure}

Participants' ratings on both \textit{Helpfulness} and \textit{Correctness} were highest for \name{} feedback (Fig. \ref{fig:writing-perceived-ratings}).

\textit{Perceived Helpfulness.}
We found a significant main effect of Feedback Type on perceived helpfulness (\textit{p} $<$ .0001). As shown in Fig. \ref{fig:writing-perceived-ratings}, \name{} feedback (\textit{M} $=$ 2.3) was rated as significantly more helpful than both Rubric-LLM (\textit{M} $=$ 1.7, \textit{p} $=$ .0016) and Text-LLM feedback (\textit{M} $=$ 1.5, \textit{p} $<$ .0001).

\textit{Perceived Correctness.}
Similarly, a significant main effect of Feedback Type on perceived correctness was observed (\textit{p} $<$ .0001). As shown in Fig. \ref{fig:writing-perceived-ratings}, \name{} feedback (\textit{M} $=$ 2.2) was rated as significantly more correct than Text-LLM feedback (\textit{M} $=$ 1.5, \textit{p} $<$ .0001) but not Rubric-LLM feedback (\textit{M} $=$ 1.7, \textit{p} $=$ n.s.).

\textit{Sense of Control.}
Despite \name{} having the highest rating, we found no significant main effect of Feedback Type on users' sense of control (see Fig. \ref{fig:writing-perceived-ratings}).

\subsubsection{Confidence (H4)}
\begin{figure}[t]
  \centering
  \includegraphics[width = 6.3cm]{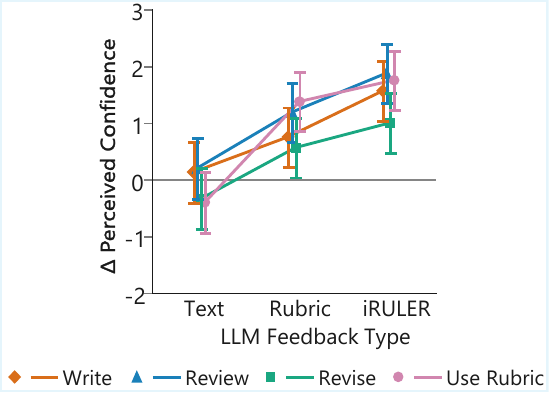}
  \caption[]{
  Change in perceived confidence ($\Delta$Confidence) for the Writing Revision experiment. \name{} yielded the largest gains, particularly for \textit{Revising} and \textit{Using Rubrics}.
  }
  \Description{A line graph showing the change in perceived confidence on the Y-axis against the three LLM Feedback Types (left to right: Text, Rubric, iRULER) on the X-axis. The Y-axis is labeled "Delta Perceived Confidence" and ranges from -2 to 3. Four line plots are shown, representing four skills: "Write" (orange line with diamonds), "Review" (blue line with triangles), "Revise" (green line with squares), and "Use Rubric" (pink line with circles).
  Generally, all four lines show an increasing trend from the Text condition to the iRULER condition.
  In the Text condition, mean values are around 0 for "Write" and "Review," and slightly below 0 for "Revise" and "Use Rubric."
  In the Rubric condition, mean values for all skills are clustered between approximately 0.5 and 1.2.
  In the iRULER condition, mean values are higher, ranging from approximately 1.0 for "Revise" to nearly 2.0 for "Review" and "Use Rubric."
  Error bars are present for all data points.}
  \label{fig:writing-perceived-confidence}
\end{figure}

Fig. \ref{fig:writing-perceived-confidence} shows the difference in user confidence ratings across Feedback Types, separately for \textit{Writing}, \textit{Reviewing}, \textit{Revising} and \textit{Rubric Use}.
\name{} significantly outperformed Text-LLM across all skills (all \textit{p} $\le$ .002) but did not differ significantly from Rubric-LLM. For \textit{Rubric Use}, Rubric-LLM was also significantly higher than Text-LLM (\textit{p} $<$ .0001).
Notably, the Text-LLM condition resulted in a decrease in confidence for some skills.


\subsubsection{Skill Transfer (H5)}
\label{sub-sub-sec:skill-transfer}
\begin{figure}[t]
  \centering
  \includegraphics[width = 4.8cm]{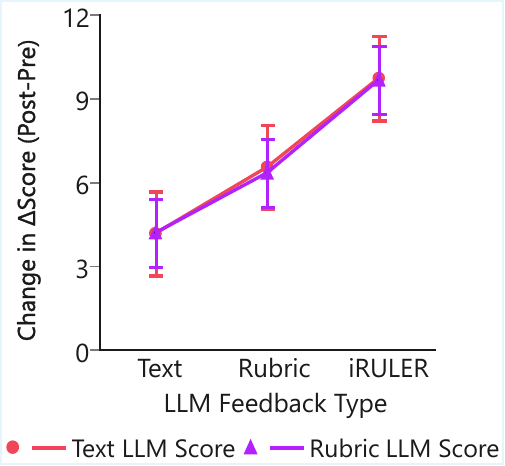}
  \caption[]{Results of skill transfer for the Writing Revision experiment, measured by the score change in $\Delta$Score (Post$-$Pre). 
  The \name{} condition leading to the greatest transferable skill gain.
  }
  \Description{A line graph showing the Change in ΔScore on the Y-axis against three LLM feedback types (left to right: Text, Rubric, iRULER) on the X-axis. The Y-axis is labeled "Change in Score (Post-Pre)" and ranges from 70 to 90 in increments of 5. 
  Two line plots, one for Text-based LLM scoring and one for Rubric-based LLM scoring, overlap significantly, depicting an increasing trend across feedback types. Across the feedback types, values are approximately 4.5 in the Text condition, approximately 6 in the Rubric condition, and approximately 9 in the iRULER condition.
  Error bars are present for all data points.}
  \label{fig:writing-skill-transfer}
\end{figure}

Fig. \ref{fig:writing-skill-transfer} shows the results for skill transfer (Change in $\Delta$Score (Post$-$Pre)).
There was a significant main effect of Feedback Type on skill transfer (\textit{p} $<$ .0001). 
Participants in \name{} improved the most (\textit{M}\textsubscript{text} $=$ 9.73, \textit{M}\textsubscript{rubric} $=$ 9.66 ) compared to Text-LLM (\textit{M}\textsubscript{text} $=$ 4.17, \textit{M}\textsubscript{rubric} $=$ 4.19, all \textit{p} $<$ .0001). 
This difference in improvement against Rubric-LLM was only significant for the Rubric-based LLM score (\textit{M}\textsubscript{rubric} $=$ 6.33, \textit{p} $=$ .0005). 


\subsection{Rubric Creation Experiment}

We recruited 36 participants (20 female, 16 male; 18--30 years old, M $=$ 23.1, SD $=$ 2.3) from our university's online platform, comprising 17 undergraduates, 19 postgraduates, and one professional certificate holder. The sample was academically diverse, with participants from fields such as Business, Engineering, and Social Sciences. 
Their familiarity with rubrics were slightly lower than for participants in the Writing Revision Experiment: 8.3\% novices, 44.4\% somewhat familiar, 41.7\% familiar, 5.6\% very familiar. 

\name{} significantly improved rubric quality (H1), perceived qualities (H3), and usage confidence (H4) compared to Text-based LLM feedback.
Non-intelligible Rubric-based LLM feedback was in-between them.
However, there was no significant difference in efficiency across Feedback Type (H2).

\subsubsection{Rubric Quality (H1)}
\name{} significantly improves overall rubric quality over Rubric-LLM and Text-LLM.

\begin{figure}[t]
  \centering
  \includegraphics[width = 5.1cm]{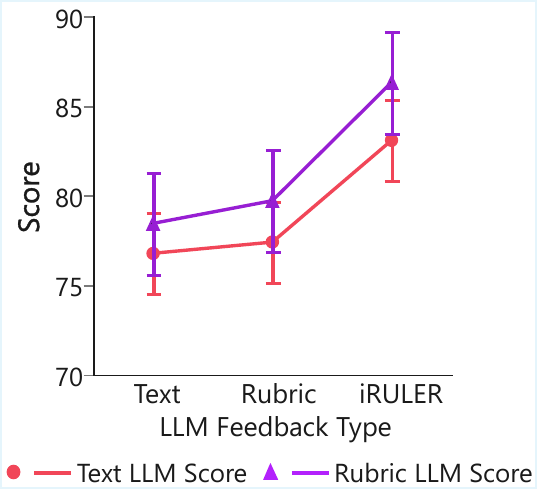}
  \caption[]{Final scores of the user-designed rubrics from Rubric Creation experiment. 
  Both methods show a consistent, significant advantage for the \name{} condition.}
  \Description{A line graph showing the final quality score of user-designed rubrics on the Y-axis against the three LLM Feedback Types (left to right: Text, Rubric, iRULER) on the X-axis. The Y-axis, labeled "Score," ranges from 70 to 90. Two line plots are shown. 
  "Rubric LLM Score" is represented by a purple line with triangles, showing mean values of 78.4 for the Text condition, 79.7 for the Rubric condition, and 86.3 for the iRULER condition. 
  "Text LLM Score" is represented by a red line with circles, showing a similar increasing trend with mean values of 76.8 for Text, 77.4 for Rubric, and 83.1 for iRULER. 
  For both line plots, the iRULER condition shows the highest mean score. Error bars are present for all data points.}
  \label{fig:rubric-score}
\end{figure}

\textit{Final Score.} 
Fig. \ref{fig:rubric-score} shows a significant effect of Feedback Type on the Text LLM score (\textit{p} $=$ .0005) and the Rubric LLM score (\textit{p} $=$ .0007). 
Participants in \name{} (\textit{M}\textsubscript{text} $=$ 83.1, \textit{M}\textsubscript{rubric} $=$ 86.3) scored significantly higher than those in Rubric-LLM (\textit{M}\textsubscript{text} $=$ 77.4, \textit{M}\textsubscript{rubric} $=$ 79.7) and Text-LLM (\textit{M}\textsubscript{text} $=$ 76.8, \textit{M}\textsubscript{rubric} $=$ 78.4, all \textit{p} $\le$ .002).

\begin{figure*}[t]
  \centering
  \includegraphics[width = 14.8cm]{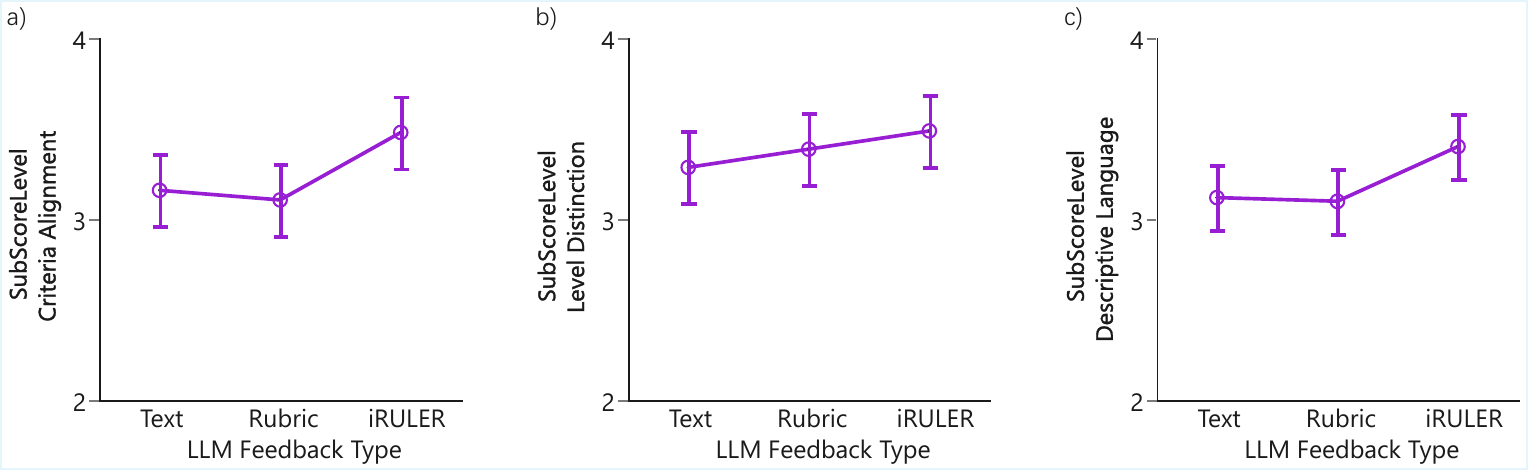}
  \caption[]{Results of the final subscore levels for the user-designed rubrics in Rubric Creation experiment. The scores for each of the three meta-rubric criteria: a) Criteria Alignment, b) Level Distinction, and c) Descriptive Language. 
  The \name{} condition led to the highest scores across all dimensions.
  }
  \Description{Three separate line graphs showing the final quality subscore levels on the Y-axis against the three LLM Feedback Types (left to right: Text, Rubric, iRULER) on the X-axis. All Y-axes range from 2 to 4 in increments of 1.
  Graph a shows "Criteria Alignment". The line is flat from Text to Rubric at approximately 3.2, and rises to approximately 3.4 for iRULER. 
  Graph b shows "Level Distinction". The line shows an increasing trend from Text to iRULER, rising from approximately 3.3 to 3.4.
  Graph c shows "Descriptive Language". The line is flat from Text to Rubric at approximately 3.2, and rises to approximately 3.3 for iRULER.
  Error bars are present for all data points in all three graphs.}
  \label{fig:rubric-subscore}
\end{figure*}

\textit{Final Subscore.} 
Fig. \ref{fig:rubric-subscore} shows the Criteria Subscores measured by the Rubric-based LLM.
We did not find any significant main effects of Feedback Type on final criteria subscores across all the criteria.


\subsubsection{Efficiency (H2)}
Unlike the writing revision task, we found no significant effect of Feedback Type on the efficiency metrics for rubric creation (Task Time and \# Iterations). 
We attribute this to the inherent complexity and exploratory nature of rubric design. 

\subsubsection{Perceived Qualities (H3)}
\begin{figure}[t]
  \centering
  \includegraphics[width = 4.8cm]{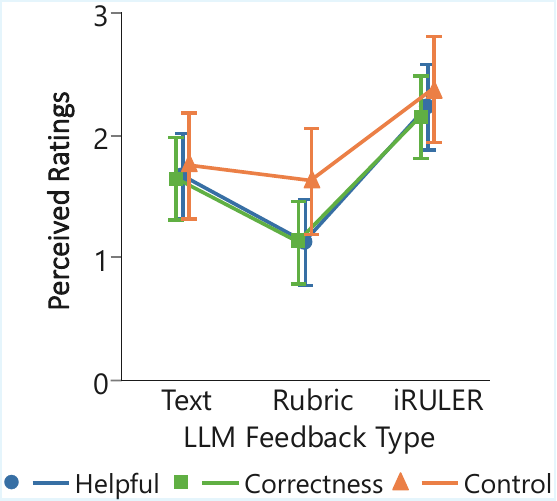}
  \caption[]{Per-trial subjective ratings for the Rubric Creation experiment. 
  \name{} was rated consistently higher than both baselines across Helpfulness, Correctness, and Control.
  }
  \Description{
  A line graph plotting "Perceived Ratings" (Y-axis, range 0 to 3) against three "LLM Feedback Types" (X-axis: Text, Rubric, iRULER). Three color-coded lines represent the metrics, showing distinct trends:
  1. Helpful (blue line, circle markers): Shows a V-shaped trend. Starts at approx 1.7 for Text, dips significantly to approx 1.1 for Rubric, and rises sharply to approx 2.2 for iRULER.
  2. Correctness (green line, square markers): Follows a similar V-shaped trend. Starts at approx 1.7 for Text, dips to approx 1.1 for Rubric, and rises to approx 2.1 for iRULER.
  3. Control (orange line, triangle markers): Remains relatively higher than the others initially. Starts at approx 1.8 for Text, dips slightly to 1.6 for Rubric, and peaks at approx 2.4 for iRULER.
  The chart highlights that the intermediate "Rubric" condition negatively impacted user ratings compared to "Text," while iRULER achieved the highest ratings across all measures.
  }
  \label{fig:rubric-result-preceived-ratings}
\end{figure}

Fig. \ref{fig:rubric-result-preceived-ratings} shows that participants' ratings on both \textit{Helpfulness} and \textit{Correctness} were highest for \name{} feedback. Notably, the Rubric-LLM condition was often rated lower than the Text-LLM baseline, suggesting that providing scores without explanation can be a frustrating user experience.

\textit{Perceived Helpfulness.}
We found a significant main effect of Feedback Type on perceived helpfulness (\textit{p} $=$ .0003). 
\name{} (\textit{M} $=$ 2.3) was rated as significantly more helpful than Rubric-LLM (\textit{M} $=$ 1.1, \textit{p} $<$ .0001) but not Text-LLM (\textit{M} $=$ 1.7, \textit{p} $=$ n.s.).

\textit{Perceived Correctness.}
Similarly, a significant main effect of Feedback Type on perceived correctness was observed (\textit{p} $=$ .0005). 
\name{} (\textit{M} $=$ 2.2) was rated as significantly more correct than Rubric-LLM (\textit{M} $=$ 1.1, \textit{p} $=$ .0001) but not Text-LLM (\textit{M} $=$ 1.7, \textit{p} $=$ n.s.).

\textit{Sense of Control.}
Similar to the writing revision task, we found no significant main effect of Feedback Type on users' sense of control.


\begin{figure}[t]
  \centering
  \includegraphics[width = 6.9cm]{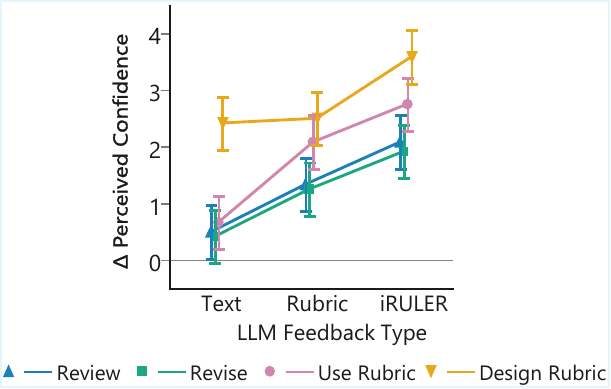}
  \caption[]{
  Change in perceived confidence ($\Delta$Confidence) for the Rubric Creation experiment. 
  \name{} resulted in the highest confidence gain compared to the baselines.
  }
  \Description{A line graph showing the change in perceived confidence on the Y-axis against the three LLM Feedback Types (left to right: Text, Rubric, iRULER) on the X-axis. The Y-axis is labeled "Delta Perceived Confidence" and ranges from -1 to 4. Four line plots are shown, representing four skills: "Review" (blue line with triangles), "Revise" (green line with squares), "Use Rubric" (pink line with circles), and "Design Rubric" (orange line with inverted triangles).
  Generally, all four lines show an increasing trend from the Text condition to the iRULER condition.
  The "Design Rubric" line is consistently positioned higher than the other three lines across all conditions, starting at a mean value of approximately 2.4 for Text and rising to approximately 3.6 for iRULER.
  The other three lines ("Review," "Revise," "Use Rubric") start clustered together near 0.5 in the Text condition, rise to a cluster between 1.2 and 2.1 in the Rubric condition, and end in a cluster between 2.0 and 2.8 in the iRULER condition.
  Error bars are present for all data points.}
  \label{fig:rubric-result-confidence}
\end{figure}


\subsubsection{Confidence (H4)}
Fig. \ref{fig:rubric-result-confidence} shows confidence gains across Feedback Types for \textit{Reviewing, Revising, Rubric Use} and \textit{Designing Rubrics}. 
\name{} yielded significantly higher gains than Text-LLM across all four skills (all \textit{p} $\le$ .0005), 
but surpassed the Rubric-LLM group only in \textit{Designing Rubrics} (\textit{p} $=$ .0012).
Similar to the Writing Revision experiment, 
Rubric-LLM significantly outperformed Text-LLM in \textit{Rubric Use} (\textit{p} $<$ .0001).


\subsection{End-to-End Experiment}

We recruited 6 participants (4 male, 2 female; 24--32 years old, \textit{M} $=$ 26.5, \textit{SD} $=$ 2.9) 
via snowball sampling. 
Regarding familiarity with rubrics, 4 reported being somewhat familiar, and 2 reported being familiar.
Participants all reported writing frequencies of at least a few times a month.

\subsubsection{Findings}
\label{sub-sec:end-to-end-findings}
Throughout both fixed and open-ended tasks, participants averaged M $=$ 1.9 revision cycles (SD $=$ 0.8) before being satisfied. 
On the 0--100 scale, mean rubric-of-rubrics and writing scores increased by 7.5 and 10.5 points respectively from the first to the final iteration.
In the open-ended task, participants freely wrote on diverse genres for application in \name{}.
Topics selected include: \textbf{science fiction}, \textbf{paper abstracts}, \textbf{technical documentation}, \textbf{social media copy} and \textbf{lateral thinking puzzles}.

\subsection{Qualitative Findings}
\label{sub-sec:qualitative-by-dg}
Having shown the effectiveness of iRULER, we performed a qualitative analysis on participant interaction and utterances in the two controlled experiments and end-to-end usage experiment.
We discuss our findings thematically, aligned with the six Design Guidelines introduced in Section~\ref{sec:design_guidelines}.
We refer to participants from the writing revision experiment (Section~\ref{sub-sec:writing-revision-eval}) as W1--W48, those from the rubric creation experiment (Section~\ref{sub-sec:rubric-creation-eval}) as R1--R36, and those from the end-to-end experiment (Section~\ref{sub-sec:end-to-end-eval}) as E1--E6.

\subsubsection{Specific}
Across systems, participants consistently sought feedback that targeted concrete elements of their work rather than general observations. 
With Text-LLM, participants noted that \textit{``it would be great if the feedback could provide more specific points, referring to specific words''} (R9), 
and that \textit{``it would be better if the tool can, like, target the specific sentence that we think can be improved... the current feedback is still very general''} (W7). 
Similarly, Rubric-based LLM users like W26 \textit{``didn't really use the rubrics much, because I feel that the rubrics in general doesn't really give specific feedback, just very general things like sentence structure could be improved' or loose organization'.''} 
In contrast, iRULER users appreciated the precision, with W47 \textit{``leaning more towards [using] it because it highlights exactly what I need to achieve whatever rubric [score].''}

\subsubsection{Scaffolded}
Participants appreciated structured guidance that helped organize their revision efforts systematically. 
Both Rubric-Based LLM and iRULER provided scaffolding through its organizational framework:
W26 systematically reflected on writing based on specific criteria, noting \textit{``[the rubric] said that I had to improve on my Content; afterwards I was consciously trying to elaborate on the Content aspect, but I couldn't improve on the Vocabulary aspect because [that] was my limit.''} 
W25 leveraged the criteria levels by examining the details within rubric cells, highlighting them with her cursor and comparing between adjacent levels to determine the next course of revisions. 
However, this scaffolding had limitations: R21 felt overwhelmed by the breadth, stating \textit{``I kind of don't really know which part I should focus on since there's so many criteria.''}

\subsubsection{Justified}
Participants also valued understanding the reasoning behind feedback scores and suggestions. 
With iRULER, participants stated that the feedback justifications contributed to a deeper understanding of revision requirements: 
\textit{``It's nice to see the explanation of why they gave that score. Because... I wouldn't have known about the overlap. I only thought about it when they mentioned it''} (R31). 
Comparing to commercially available software, W39 found iRULER superior: \textit{``I used to use QuillBot or Grammarly. The operation behind it is very vague. I don't know why it gave me a score. But it gave me a change. With [iRULER], I think it is an improvement in learning. I also know what goal I want to achieve.''}

\subsubsection{Actionable}
Participants generally needed feedback that clearly indicated what changes to make. Those in the Text-LLM condition found the feedback helpful initially, but eventually sought more actionable guidance. 
W11 had a \textit{``frustrating experience''} because \textit{``when I made changes to the essay, ... a lot of the time, the feedback remained static and the score also did not change much.''} 
W13 felt that \textit{``even after changes were made, there was no change in the score, which made me want to give up,''} and R10 thought that \textit{``after I generated my evaluation of performance, [the system] didn't consider those points that I added.''}

In contrast, participants consistently found iRULER feedback actionable in various avenues such as generating ideas (W38, R28, R32), improving structure (W47, R36), and suggesting new phrases (W36, W44, W45). 
W47 explicitly noted that \textit{``before I used [the counterfactual actions], I was a bit frustrated because the rubric did not exactly tell me what to improve on.''} 
However, the increased actionability also revealed a problem of LLM verbosity: W48 felt the system \textit{``over-generate[s] in a sense. You make it too [sophisticated]... I feel it might just keep generating longer and longer texts, but there's no content that's being added.''} 
R28 echoed this in the rubric task: \textit{``in order to correct its shortcomings, [the AI revision] says more... so it feels like it adds more work, because I have to read it again---the long text after the change---and then shorten it again.''}

\subsubsection{Qualified}

In the rubric creation experiment, participants found the meta-rubric useful in qualifying their own rubrics. 
R15 thought \textit{``it's quite clear, because the three things that it measures, Criteria, Level Distinction, and Language... these are the key aspects of what makes a rubric a rubric, so it's also quite clear what could improve.''} 
She also felt that \textit{``the overall score is very helpful with the color background. It's very obvious, like where you stand... So, you immediately know without looking at the descriptors, the overall quality of the task.''}
R14 found specific meta-rubric criteria useful: \textit{``if you look at Criteria Alignment... it evaluates if [my] 3 criteria can cover all aspects of the copy.''}

In the end-to-end experiment, participants found that applying their rubrics to concrete examples helped them qualify whether their criteria were aligned with their intentions.
E2 could \textit{``see whether the sample text is being graded correctly and accurately according to my own rubrics.''} 
E4 remarked that \textit{``having me write a rubric from scratch and then immediately use it for grading helps me spot flaws in my rubric right away, which is incredibly valuable.''}

Moreover, participants 
appreciated the consistency in both \textit{writing revision} and \textit{rubric creation} UIs. 
E3 felt that \textit{``the interfaces for rubric and writing are very consistent. Using my rubric to assess writing feels very intuitive because I'm already familiar with it.''} 

\subsubsection{Refinable}
In the end-to-end experiment, participants iteratively and incrementally refined their rubrics by looping between the rubric editing and writing revision tasks.
E5 felt that \textit{``creating my own rubric before using it for grading helps me continually clarify exactly what aspects I want to evaluate and the distinctions between each level,''}
Consequently, he felt \textit{``a greater sense of involvement compared to simply using a pre-made rubric.''} 
E1 \textit{``realized like maybe Organization contains Language as well. Like it tries to edit grammar as well. So I want to make this more specific to just structure rather than including the [grammar]''}. 
Lamenting about his daily work, E2 remarked that using iRULER \textit{``really hammers in how boring a teacher's job is''}, and that iRULER \textit{``definitely saves a lot of time''}.

Despite the opportunity for refinement, some participants found this mentally demanding.
E4 felt that \textit{``the information overload was overwhelming---trying to think about both `Is this rubric well-written?' and `Is this essay well-written?' at the same time felt like my brain was splitting in two''}.
E5 struggled with \textit{``switching back and forth between the perspective of a `teacher' and an `editor' within a single session makes me feel a bit burdened by the role switching.''}






\subsection{Human Expert Validation of LLM-as-a-Judge}


We recruited two expert raters (male; ages 32, 35) who were English writing tutors at a university language tutoring center.
They had 5 and 8 years' experience of teaching and using rubric-based assessment and scoring essays.
Table~\ref{tab:expert_llm_agreement} summarizes the results.

\begin{table}[t]
\centering
\small
\setlength{\tabcolsep}{4pt}

\caption[]{
    Human expert validation of scoring metrics for Rubric-LLM and Text-LLM. 
}
\label{tab:expert_llm_agreement}
\begin{tabular}{lcccccc}
\toprule
& \multicolumn{2}{c}{\makecell{Inter-Rater\\Reliability}} 
& \multicolumn{2}{c}{\makecell{Rubric-LLM\\vs. Experts}} 
& \multicolumn{2}{c}{\makecell{Text-LLM\\vs. Experts}} \\
\cmidrule(r){2-3} \cmidrule(lr){4-5} \cmidrule(l){6-7}
Dimension      & $\alpha$ & 95\% CI   & QWK & 95\% CI   & QWK & 95\% CI   \\
\midrule
Content        & 0.80 & 0.55--0.95 & 0.53 & 0.35--0.69 & --   & --        \\
Organization   & 0.82 & 0.58--0.96 & 0.58 & 0.44--0.70 & --   & --        \\
Vocabulary     & 0.85 & 0.61--1.00 & 0.66 & 0.51--0.79 & --   & --        \\
Language Use   & 0.87 & 0.67--1.00 & 0.73 & 0.61--0.83 & --   & --        \\
\midrule
Total Score    & 0.96 & 0.93--0.98 & 0.88 & 0.81--0.92 & 0.76 & 0.67--0.81 \\
\bottomrule
\end{tabular}
\end{table}

\subsubsection{Expert Inter-Rater Reliability}

Both experts strongly agreed with each other for all criteria subscores ($\alpha \ge .80$).
Their agreement on the Total Score was even higher ($\alpha = .96$).

\subsubsection{Expert-LLM Score Agreement}

The alignment between the experts and the Rubric-LLM total score was high (QWK $= .88$), and higher than the moderately high alignment with Text-LLM (QWK $= .76$).
Expert alignment with criteria subscores were within moderate to substantial levels ($.53$--$.73$)~\cite{landis1977application}.
\section{Discussion}
Having validated \name{}'s effectiveness in improving reviewing and revision, we now turn to the broader implications for theory and design in human-AI collaboration.

\subsection{Providing Specified, Scaffolded, Justified, Actionable and Qualified Feedback with iRULER}
A core contribution of our work is a system that streamlines the often-amorphous process of offering high-quality feedback~\cite{Price01052010, dawson2019makes, ajjawi2017researching} by translating abstract guidelines into an interactive workflow. 

The result of \name{} helping users produce higher-quality artifacts (H1) demonstrates the effectiveness of feedback that is \textit{specific}, \textit{scaffolded}, \textit{justified} and \textit{actionable}. We achieve this by treating the rubric as a shared reference, anchoring the LLM's evaluation in an explicit structure, and turning an open-ended evaluation into a constrained, evidence-based judgment. The high ratings for helpfulness and correctness (H3) further suggest that \textit{justified} and \textit{specific} rubric criteria are vital for building user trust.
Furthermore, our results highlight how this structured interaction can be made effective. Significant gains in user agency (H3) and confidence (H4) emphasize the importance of a well-\textit{scaffolded} user interface. The rubric provides a structured framework to empower users rather than overwhelm them.


Finally, by recursively applying DG1--4, the Rubric Creation experiment validates our fifth guideline: feedback must be \textit{qualified}. This meta-level evaluation establishes a metacognitive loop, prompting reflection on both the artifact and the criteria to uphold collaborative integrity. The performance results (H1) demonstrate the significant value of this approach.

\subsection{Toward Personalized Feedback with Refinable User-Defined Criteria}

Human-centered AI prioritizes assistance aligned with users' unique goals and context~\cite{amershi2019guidelines, yang2020re, wadhwa2025evalagent}. 
Moving beyond \textit{implicit} personalization, where models infer preferences from behavioral data~\cite{pavlik2013review,shen2005implicit}, \name{} enables \emph{explicit} personalization through user-defined criteria~\cite{li2025beyond} turning rubrics from static scoring tools into dynamic, refinable mechanisms for steering AI behavior.

Our findings show that such personalization is effective when user-defined criteria are 
both \emph{qualified} and \emph{refinable}. 
In our end-to-end experiment, the application of rubrics to concrete writing samples served as a verification step.
This allowed participants to audit the model's performance against their internal standards and ``debug'' their criteria in real time.
Through this process, participants were able to identify logical flaws in their rubrics and sharpen level descriptions, efficiently iterating towards their own specific needs. 
This mirrors recent observations of \emph{criteria drift}, where evaluators naturally add, split, and revise criteria over time~\cite{shankar2024validates}.
Consequently, our results extend this perspective to rubric-based revision: effective personalization requires supporting users in iteratively articulating, qualifying, and refining their evaluative criteria, rather than simply adapting model outputs.

\subsection{Navigating Design Tensions in Collaborative AI}
\subsubsection{Balancing Guidance and Agency}

A general tension in collaborative AI, spanning coding, design and writing, lies in balancing effective guidance against the risk of over-reliance and de-skilling~\cite{kaur2020interpreting, passi2022overreliance}.
In our study, Text-LLM participants received generic feedback and iterated the most, yet reported a decrease in revision confidence.


Our results showed that \name{} addresses this through our design guidelines. 
Grounding feedback in \textit{specific}, \textit{scaffolded} criteria provides structural guardrails, anchoring interactions in explicit goals.
\textit{Justified} \textit{Why} and \textit{Why-Not} explanations elicit interactive interpretation of feedback, fostering deeper understanding~\cite{chi2014icap}.
\textit{Actionable} \textit{How-To} examples are presented as suggestions that require user review and selection~\cite{chi2014icap, laban2024beyond}.
While meta-level feedback \textit{qualifies} the rubric, continued iteration on real writing helps users internalize and \textit{refine} their criteria. 
The skill transfer effects also suggest that this more intelligible form of assistance can support learning instead of short-circuiting it. 
By exposing the evaluation logic, \name{} enables a kind of \textit{forward simulation}~\cite{bo2024incremental, hase2020evaluating}, helping users anticipate how drafts will be evaluated.
This aligns with work showing that active engagement with feedback encourages reflection~\cite{choi2023benefit, sargeant2009reflection} and improves learning outcomes~\cite{chi2014icap}. 
Taken together, these insights are promising for a \textit{learning-oriented} form of co-creation~\cite{disalvo2016participatory,nicholson2022participatory}: users are not only revising with the AI, but also tuning the evaluative lens that judges their artifact.


\subsubsection{Balancing Generative Expressiveness and Structured Constraints}
Another tension lies in balancing generative freedom with the need for concise, focused guidance~\cite{ippolito2022creative, lee2022coauthor}.
Although the model could identify issues, without constraints, LLMs may prioritize statistical probability over utility~\cite{holtzman2019curious}, leading to the ``overly verbose'' or ``rambling'' tendencies observed by our participants.

\name{} addresses this by using the rubric as a structured constraint on the model's vast output space. 
By forcing the AI to ground every piece of feedback in \textit{specific} criteria and performance levels, the rubric acts as a \textit{scaffolding} tool, steering the model's generation towards what is most relevant to users. 
The \textit{Why}/\textit{Why Not}/\textit{How To} structure further constrains output into \textit{justified} and \textit{actionable} formats. 
This suggests that for co-creative tools to be effective, they require not just powerful generative engines, but also user-defined structures that channel that power effectively.

\subsubsection{Writing-Specific: Balancing Standardization and Authorial Voice}
Writing presents a unique challenge: a trade-off between structural correctness and subjective expression~\cite{jonsson2007use,ippolito2022creative}. 
A rigid, automated evaluator risks ``homogenizing'' text, stripping away unique voice in favor of statistical probability, or encouraging ``reward hacking'' where users optimize for the AI's hidden preferences rather than communicative goals~\cite{shankar2024validates}.

In our end-to-end experiment, this tension manifested when participants felt the AI's suggestions failed to capture subtle criteria such as ``Creativity''. 
\name{} could mitigate this risk through \textit{refinability}. 
By treating the rubric as a user-defined, evolving artifact rather than a fixed standard, \name{} allows writers to codify their own definition of quality. 
This helps align the AI's \textit{justified} critiques with the user's intended voice, attenuating diminished creativity while still offering structured scaffolding.

\subsubsection{Writing-Specific: Balancing Surface Polish and Rhetorical Structure}
Another tension in writing support is balancing surface-level polishing and deep, structural revision. 
Writing permits ``polished incoherence'', text that is grammatically well-formed on the surface, yet rhetorically weak underneath~\cite{jakesch2023co}. 
Generative models excel at locally enhancing fluency~\cite{lee2022coauthor}, potentially inducing ``premature satisfaction'' where users overlook underlying argumentative gaps due to the smoothness of AI-generated prose~\cite{gero2022sparks, zhang2023visar}.

\name{} could address this by decoupling evaluation criteria with rubrics. 
Unlike conversational interfaces that often conflate stylistic edits with substantive content suggestions~\cite{laban2024beyond,zhang2023visar}, 
\textit{specific} and \textit{scaffolded} rubrics can isolate \textit{Organization} and \textit{Content} from \textit{Language Use}. 
This design successfully directed user attention toward deeper rhetorical improvements, as evidenced by the higher gains in \textit{Organization} subscores compared to baselines,
and motivated participants to \textit{refine} initially broad criteria that otherwise biased the model toward surface-level editing.

\subsection{Toward a Generalizable Framework for Intelligible Co-Evaluation}
We propose that \name{}'s core contribution is a shared, user-defined structure that mediates the evaluation of generative outputs, offering a \textit{potentially} generalizable framework for human-AI collaboration.
Having evaluated with argumentative writing in our writing revision experiment, social media ad copy and public speaking in our rubric creation experiment, and a diverse range of open-ended objective and subjective topics (science fiction, paper abstracts, technical documentation, social media copy and lateral thinking puzzles) in our qualitative end-to-end experiment, 
our results demonstrates the framework's broad applicability across both objective and subjective writing domains.

The generalizability of our framework lies in the inherent customizability of rubrics.
Unlike argumentative writing where success metrics are relatively objective~\cite{panadero2013rubrics}, highly subjective tasks like creative writing, graphic design, or music composition often have flexible, evolving, and even conflicting goals~\cite{ippolito2022creative,frich2019mapping}.
Our end-to-end experiment revealed that the \textit{refinability} design guideline can help to manage this ambiguity 
by letting users iteratively refine criteria as their vision evolves, 
and by supporting high-frequency ``task-switching'' supported by our two UIs.


However, we acknowledge that not all criteria are amenable to purely LLM-based linguistic analysis. 
Veracity as a criterion cannot be evaluated via language patterns alone~\cite{liu2023trustworthy}; 
determining whether a claim and its supporting statements are factually true requires external verification capabilities, such as connections to Knowledge Graphs or Retrieval-Augmented Generation pipelines~\cite{zhang2025siren, huang2025survey}. 
Stylistic judgments that depend heavily on taste, such as distinguishing between humorous/sarcastic and cordial/innocuous tones, also remain challenging for general purpose models~\cite{song2025large}.
These subjective preferences may require users to explicitly anchor these definitions first, to align models with their specific intent.


Beyond writing tasks, structured evaluation is already a standard practice, from design critiques~\cite{haraguchi2024can} to assessing music composition~\cite{chen2021attend}, suggesting potential for \name{} to expand.
However, non-verbal concepts such as \textit{visual balance} in design or \textit{timbre} in audio are inherently difficult for an LLM to interpret through text alone.
Generalizing to these modalities would require integrating multi-modal LLMs (MLLM) capable of grounding textual criteria in visual or auditory inputs~\cite{lin2024designprobe, liu2024mumu}, which may pose issues with data sparsity and out-of-distribution occurrences~\cite{zhang2024out, wu2025f}.
We envision a paradigm where users provide definitions alongside multi-modal few-shot examples of `good' and `bad' adherence to a criterion. The system then leverages in-context learning~\cite{brown2020language} to align its representations with the user's tacit knowledge.



Finally, we recognize that criteria-based rubrics represent only one approach for evaluation.
In composition pedagogy, \textit{rhetorical frameworks} emphasize the dynamic interplay between exigence, audience, and constraints~\cite{bitzer1968rhetorical, zhao2025plan}, which rubrics may struggle to capture holistically.
Complementary to this are structural techniques like \textit{reverse outlining}, where writers abstract the main outline of a draft to evaluate the broader logical flow \cite{flower1989problem}.
Future work could leverage LLMs to adopt these alternative frameworks for comprehensive rhetorical and structural evaluations.

\subsection{Limitations and Future Work}
\label{sub-sec:validity}

Several limitations highlight directions for future research. 
First, while decoupling writing revision and rubric creation was necessary to isolate task-specific effects and deconfound variables, this design limits our quantitative understanding of the integrated, recursive workflow.
Although our preliminary end-to-end experiment provided meaningful qualitative support, 
future work requires large-scale quantitative validation involving diverse stakeholders.

Secondly, our writing revision experiment demonstrated significant immediate skill transfer but was limited in assessing long-term retention.
Longitudinal deployments are required to examine 
whether these benefits persist.

Finally, 
while the rubric creation and end-to-end experiments covered diverse topics and offered initial insights into the system's broader applicability,
our controlled writing revision experiment focused mainly on argumentative writing, 
a widely used genre with well-established improvement guidelines  
and the availability of expert-scored datasets (ICNALE~\cite{ishikawa2018icnale}), 
which provided a robust baseline for validation.
Future work should systematically assess \name{}'s generalizability to highly subjective or multi-modal domains, such as creative writing and UI design, which pose unique challenges for generating actionable counterfactuals~\cite{chakrabarty2024art, o2015designscape}.

\section{Conclusion}
We presented six design guidelines for user-defined feedback, operationalized through \name{}, an interactive system for rubric-based human-AI co-evaluation. 
By recursively applying feedback to both writing and rubrics, our dual-level framework offers a unique paradigm for human-AI co-evaluation. 
Empirical evaluations confirm its efficacy, with users rating \name{} as more intelligible, trustworthy, and empowering than traditional baselines. 
Beyond technical contributions, \name{} advances the goal of human-centered AI, shifting the role of LLMs from opaque graders to transparent, collaborative partners. 
These findings highlight the promise of principled, user-driven tools in transforming education and creative workflows, driving both superior results and deeper learning.

\begin{acks}
We thank Ayrton San Joaquin and Yehoon Ahn for their contributions to the early ideation and discussions of this work.
This research is supported by the National Research Foundation, Singapore and Infocomm Media Development Authority (Award No. DTC-RGC-09), the Ministry of Education, Singapore (Award No. MOE-T2EP20121-0010), the NUS Institute for Health Innovation \& Technology (iHealthtech), the DesCartes programme supported by the National Research Foundation, Prime Minister's Office, Singapore under its Campus for Research Excellence and Technological Enterprise (CREATE) programme, and the AI Interdisciplinary Institute ANITI under the Grant agreement n\textsuperscript{o}ANR-23-IACL-0002.
Any opinions, findings and conclusions or recommendations expressed in this material are those of the author(s) and do not reflect the views of the National Research Foundation, Singapore and Infocomm Media Development Authority.

\end{acks}

\bibliographystyle{ACM-Reference-Format}

\bibliography{main}

\clearpage

\appendix

\section{LLM Prompts}
\label{appendix:prompts}

This appendix provides the prompts used by \name{} and baseline systems. 
The Model is \textbf{OpenAI's GPT-4.1}, the temperature was varied for each specific task to balance creativity and consistency.
Dynamic content, which is programmatically inserted by the client-side application, is indicated by [-- text in brackets --].

\subsection{Prompts in iRULER}
\label{appendix:iruler-prompts}

\subsubsection{Evaluation Prompts}

\label{appendix:iruler-evaluation-prompts}
This prompt uses a unified template across the Writing Revision and Rubric Creation. And is called for each individual criterion in a rubric. It instructs the LLM to generate a score, a \textit{Why} explanation for the chosen score, and a \textit{Why Not} explanation for all other score levels.
\textit{Temperature: 0}.

\begin{quote}
\scriptsize
\setlength{\parskip}{0pt} 
\setlength{\baselineskip}{6pt}

\texttt{\textbf{<USER PROMPT>}} \\
\texttt{<role, task>}\\
\texttt{You are an expert in evaluating <user artifact> in the dimension of [-- A placeholder for the current meta-criterion's name --]. You should evaluate the <user rubric> strictly based on the <Criteria>, and give a detailed reason for your evaluation. Return a json object...}\\
\texttt{</role, task>}\\
\\
\texttt{<task description>}\\
\texttt{[-- A placeholder for the overall task description --]}\\
\texttt{</task description>}\\
\\
\texttt{<user artifact>}\\
\texttt{[-- A placeholder for the user's current artifact --]}\\
\texttt{</user artifact>}\\
\\
\texttt{<Criteria>}\\
\texttt{[-- A JSON object representing the specific criterion and its level descriptors --]}\\
\texttt{</Criteria>}\\
\\
\texttt{<instructions>}\\
\texttt{  You have to do the following 7 steps:\\
  1. You need understand the <Evaluation Criteria>. Which has a score scale from score\_[-- score --] to score\_1, under each score, there is a text field which indicates the requirements of getting the corresponding score. Typically, the higher the score, the better the user rubric in this criteria. [-- task description --].\\
  2. For each Criteria, your evaluation for the <user rubric> should strictly follow the criteria, and for each of those evaluation criteria, rigorously match the <user rubric> with the score\_[-- score --] to score\_1 of the text attributes and select the score that best matches the criteria and set its 'checked' field to true. Only one score can be true in each object of the array.\\
  3. Adopt a moderate evaluation approach: prioritize fairness by recognizing strengths in the rubric, avoid overly harsh penalties for minor errors, and focus on the overall alignment with the criteria.\\
  4. For the score marked with "checked": true, provide a detailed "why" explanation with concrete examples. Write this in its "reason" field, explaining why this score was selected over the others.\\
  5. For each score marked with "checked": false, provide a detailed "why not" explanation with concrete examples. Write this in its "reason" field, explaining why the score was not selected and give examples of what the user rubric did not meet the criteria.\\
  6. Use an Overall–Supporting structure in all "reason" fields. Start with a concise overall evaluation of the score (1–2 sentences), followed by a bulleted list of specific reasons and concrete examples that support the main point. The "reason" field should be written in Markdown format with line breaks, not as a list.\\
  7. Format your response as a single JSON object (see example below). Do not include any other text outside the JSON.}\\
\texttt{</instructions>}\\
\\
\texttt{<output format example (In json format)>}\\
\texttt{[-- A JSON structure showing the criterion with "checked" and "reason" fields for each score level --]}\\
\texttt{</output format example>}
\end{quote}

\subsubsection{Counterfactual Writing Generation Prompt (How-to)}
\label{appendix:how-to-writing-prompt}
This prompt is triggered when a user requests a \textit{How To} example for current writing based on the criteria, current score and selected level. It is a chained prompt that includes the context from the previous evaluation and enforces the principle of minimal modification.
\textit{Temperature: 0}.

\begin{quote}
\scriptsize
\setlength{\baselineskip}{6pt} 
\setlength{\parskip}{0pt} 
\texttt{\textbf{<USER PROMPT>}} \\
\texttt{<role, task>}\\
\texttt{As an AI writing expert, you will revise a given text, <user writing>, to meet a specific score on a rubric, <rubric>. Your goal is to function as a precise and logical revision tool... The revision can improve or reduce the user writing based on the rubric, but you should make the fewest changes possible...}\\
\texttt{</role, task>}\\
\\
\texttt{<Core Principle: Minimal Modification>}\\
\texttt{Your primary directive is to adhere to the principle of  minimal modification. You must make the fewest changes possible to achieve the target score. For each potential change, ask yourself: "Is this modification absolutely necessary to meet the criteria of the target score?" If the original text already meets a criterion, do not change it...}\\
\texttt{</Core Principle: Minimal Modification>}\\
\\
\texttt{<user writing>}\\
\texttt{[-- A placeholder for the user's current writing artifact --]}\\
\texttt{</user writing>}\\
\\
\texttt{<rubric, current score, target score and rationale>}\\
\texttt{[-- Programmatically generated text describing the rubric, the current score, the target score, and the "Why Not" explanation from the previous step --]}\\
\texttt{</rubric, current score, target score and rationale>}\\
\\
\texttt{<instructions>}\\
\texttt{1. \textbf{Analyze the Rubric}: Thoroughly analyze the <rubric> to understand the criteria... Focus on the specific language that distinguishes the current score from the target score...}\\
\texttt{2. \textbf{Revise the Writing}: Revise the <user\_writing>... strictly adhering to the <rubric> and the  Core Principle of Minimal Modification ... Prioritize small adjustments... Preserve the original style...}\\
\texttt{3. \textbf{Provide Detailed Rationale}: For each modified sentence, include a justification... Each justification must contain: 'sentence' and 'reason'... The 'reason' must explicitly tie the change to the rubric’s language...}\\
\texttt{4. \textbf{Format the Output}: Return a single JSON object... Do not include any text or explanations outside the JSON structure.}\\
\texttt{</instructions>}\\
\\
\texttt{<output format example>}\\
\texttt{\{}\\
\texttt{  "revised\_writing": "The revised <user writing>...",}\\
\texttt{  "reason": [\{ "sentence": "...", "reason": "..." \}, ...]}\\
\texttt{\}}\\
\texttt{</output format example>}
\end{quote}

\subsubsection{Counterfactual Rubric Generation Prompt (How-to):}
\label{appendix:how-to-rubric-prompt}
This prompt is triggered when a user requests a \textit{How To} example for current rubric based on the criteria, current score and selected level. It is a chained prompt that includes the context from the previous evaluation to guide a comprehensive revision of the entire user rubric. 
\textit{Temperature: 0}.

\begin{quote}
\scriptsize
\setlength{\parskip}{0pt} 
\setlength{\baselineskip}{6pt} 
\texttt{\textbf{<USER PROMPT>}} \\
\texttt{<role, task>}\\
\texttt{You are an expert rubric designer with extensive experience in developing high-quality assessment rubrics. Your task is to comprehensively revise the provided <user rubric> based on meta-rubric evaluation feedback to achieve a higher quality standard.
You will receive: (1) a user-created rubric, (2) meta-rubric evaluation results showing current performance and target performance, and (3) detailed rationale explaining gaps between current and target scores.
Your goal is to systematically improve the entire rubric structure to meet the target quality level.}\\
\texttt{</role, task>}\\
\\
\texttt{<user rubric>}\\
\texttt{[-- The full JSON object of the user-designed rubric --]}\\
\texttt{</user rubric>}\\
\\
\texttt{<meta-rubric evaluation feedback>}\\
\texttt{[-- Programmatically generated text describing the meta-rubric, the current score, the target score, and the "Why Not" explanation from the previous step --]}\\
\texttt{</meta-rubric evaluation feedback>}\\
\\
\texttt{<revision principles>}\\
\texttt{1. Comprehensive Enhancement: Address all identified weaknesses across the entire rubric, not just individual score levels.\\
2. Systematic Improvement: Ensure all score levels within each criterion work together as a coherent progression.\\
3. Scalable Design: Adapt your revisions to work with any number of score levels (the rubric may have 3--6 score levels).\\
4. Quality Standards: Focus on achieving the target score by addressing the specific gaps identified in the <meta-rubric evaluation feedback>.\\
5. Consistency: Maintain consistent language, structure, and expectations across all dimensions and score levels.}\\
\texttt{</revision principles>}\\
\\
\texttt{<instructions>}\\
\texttt{1. Analyze the <meta-rubric evaluation feedback>: Carefully examine the meta-rubric evaluation to understand: What the current score represents and why it was assigned. What the target score requires and why it wasn't achieved. Specific areas needing improvement based on the detailed rationale\\
2. Identify Revision Areas: Based on the <meta-rubric evaluation feedback>, determine what aspects of the rubric need enhancement.\\
3. Comprehensive Revision: Systematically revise the entire rubric structure:\\
Improve descriptions across all score levels to create clear, logical progressions. Enhance clarity and distinctiveness of each criterion. Refine language for precision, consistency, and professionalism. Ensure all elements work together effectively. Strengthen connections to the specific assignment context when applicable\\
4. Quality Assurance: Ensure your revised rubric: Addresses all issues identified in the meta-rubric <meta-rubric evaluation feedback>. Maintains the original rubric structure (dimensions, percentages, score scale). Creates clear distinctions between all performance levels. Demonstrates the improvements needed to achieve the target score\\
5. Output Formatting: Return a complete JSON object matching the exact structure of the input rubric, with all score levels revised as needed.}\\
\texttt{</instructions>}\\
\\
\texttt{<output format>}\\
\texttt{Return the complete revised rubric as a JSON array maintaining the exact same structure as the <user rubric>, with all necessary improvements applied across all score levels. The output should preserve: All dimension names and descriptions (unless they need improvement based on feedback). All percentage weightings. All existing score level keys (score\_1, score\_2, score\_3, etc.). The "checked": false property for all score levels. Dynamically adapt to the number of score levels present in the input rubric. Your output structure should exactly match the input structure.}\\
\texttt{</output format>}
\end{quote}

\subsection{Prompts in Baseline Systems}
\label{appendix:baseline-prompts}


\subsubsection{Text-LLM Prompt in Writing Revision}
\label{appendix:text-llm-writing-prompt}
This prompt is used to generate a holistic score and a brief textual comment, simulating a standard, non-rubric-based AI feedback system.
\textit{Temperature: 0}.

\begin{quote}
\scriptsize
\setlength{\baselineskip}{6pt} 
\setlength{\parskip}{0pt} 
\texttt{\textbf{<SYSTEM PROMPT>}} \\
\texttt{<role, task>}\\
\texttt{You are an expert writing evaluator with extensive experience in assessing academic and professional writing, specifically for ESL (English as a Second Language) learners. Your task is to provide a comprehensive yet concise holistic assessment of the given writing by applying the ESL Composition Profile.}\\
\texttt{</role, task>}\\
\\
\texttt{<evaluation methodology>}\\
\texttt{You should follow a systematic approach to evaluate the writing:}\\
\texttt{1. \textbf{Content Analysis}: First, analyze the writing to understand its purpose and substance.}\\
\texttt{2. \textbf{Evaluation using ESL Composition Profile}: Evaluate the writing against the five core dimensions of the ESL Composition Profile: Content, Organization, Vocabulary, and Language Use.}\\
\texttt{3. \textbf{Systematic Assessment}: Evaluate the writing against each dimension on a 4-point scale.}\\
\texttt{4. \textbf{Holistic Scoring}: Synthesize your dimensional assessments into a single holistic score.}\\
\texttt{5. \textbf{Synthesized Feedback}: Identify the greatest strength and the 1--2 most critical areas for improvement, then provide concise, actionable feedback.}\\
\texttt{</evaluation methodology>}\\
\\
\texttt{<instructions>}\\
\texttt{You have to do the following 6 steps:}\\
\texttt{1. \textbf{Analyze the Writing}: Examine the writing to understand its main ideas, purpose, and overall message.}\\
\texttt{2. \textbf{Evaluate with ESL Composition Profile}: Instead of creating a rubric, you must evaluate the writing based on the four fixed dimensions of the ESL Composition Profile:
   -  Content : The substance of the writing, the development of the topic, the relevance and clarity of ideas.
   -  Organization : The logical structure, coherence, paragraphing, and use of transitions.
   -  Vocabulary : The range, precision, and appropriateness of word choice.
   -  Language Use : The effectiveness and correctness of sentence structure, grammar, and idiomatic expressions.}\\
\texttt{3. \textbf{Dimensional Assessment}:  For each of the four dimensions, mentally evaluate the writing on a 4-point scale:
   - 4: Excellent/Exceeds expectations
   - 3: Good/Meets expectations
   - 2: Satisfactory/Partially meets expectations
   - 1: Needs improvement/Below expectations}\\
\texttt{4. \textbf{Holistic Integration}: Combine your dimensional scores into a single holistic score from 0--100, considering the relative importance of each dimension for this type of writing.}\\
\texttt{5. \textbf{Synthesize Feedback}: Based on your dimensional assessment, identify the single greatest strength and the one or two most critical areas for improvement. Synthesize this into a concise, actionable feedback paragraph (2--3 sentences). The feedback should first acknowledge the strength and then clearly state the areas needing improvement, referencing the relevant ESL Composition Profile dimensions (e.g., Content, Organization, etc.).}\\
\texttt{6. \textbf{Quality Assurance}: Ensure your score reflects the actual quality demonstrated and your feedback is encouraging yet honest.}\\
\texttt{</instructions>}\\
\\
\texttt{<output format>}\\
\texttt{Return ONLY a JSON object in this exact format:}\\
\texttt{\{}\\
\texttt{  "score": [whole number from 0--100],}\\
\texttt{  "comment": "[A concise paragraph of 2--3 sentences...]''} \\
\texttt{\}} \\
\texttt{</output format>}\\
\\
\texttt{\textbf{<USER PROMPT>}} \\
\texttt{<user writing>}\\
\texttt{[-- A placeholder for the user's current writing artifact --]}\\
\texttt{</user writing>}\\
\\
\texttt{Please evaluate this writing following the systematic methodology described above.}
\end{quote}

\subsubsection{Text-LLM Prompt in Rubric Creation}
\label{appendix:text-llm-rubric-prompt}

\label{prompt:rubric-text-llm}
This prompt is used to generate a holistic score and a brief textual comment for a user-designed rubric, simulating a standard, non-interactive AI evaluation. The full text of the meta-rubric is included in the system prompt to guide the LLM's judgment. 
\textit{Temperature: 0}.

\begin{quote}
\scriptsize
\setlength{\baselineskip}{6pt} 
\setlength{\parskip}{0pt} 
\texttt{\textbf{<SYSTEM PROMPT>}} \\
\texttt{<role, task>}\\
\texttt{You are an expert in rubric design and assessment... Your task is to provide a comprehensive assessment of the given rubric based on established meta-rubric criteria...}\\
\texttt{</role, task>}\\
\\
\texttt{<evaluation methodology>}\\
\texttt{You should follow a systematic approach to evaluate the rubric quality: 1. Content Analysis... 2. Evaluation using Meta-Rubric Standards... 3. Systematic Assessment... 4. Holistic Scoring... 5. Synthesized Feedback.}\\
\texttt{</evaluation methodology>}\\
\\
\texttt{<meta-rubric standards>}\\
\texttt{Evaluate the rubric based on these three core dimensions:}\\
\texttt{1. \textbf{Criteria Alignment}: Score 4 (All criteria measure...), Score 3 (Most criteria...), Score 2 (Some criteria...), Score 1 (Few or no criteria...).}\\
\texttt{2. \textbf{Level Distinction}: Score 4 (The scale creates a clear, logical...), Score 3 (The scale shows a logical...), Score 2 (The progression is sometimes...), Score 1 (The progression is illogical...).}\\
\texttt{3. \textbf{Descriptive Language}: Score 4 (Language describes precise...), Score 3 (Language is mostly descriptive...), Score 2 (Language mixes descriptive...), Score 1 (Language is primarily subjective...).}\\
\texttt{</meta-rubric standards>}\\
\\
\texttt{<instructions>}\\
\texttt{You have to do the following 6 steps: \\
1. Analyze the Rubric Structure: Examine the rubric to understand its dimensions, scoring scale, criteria descriptions, and weighting.\\
2. Evaluate with Meta-Rubric Standards: Assess the rubric against each of the three meta-rubric dimensions listed above.\\
3. Dimensional Assessment: For each of the three meta-dimensions, mentally evaluate the rubric on a 4-point scale: 4: Excellent/Exceeds standards. 3: Good/Meets standards. 2: Satisfactory/Partially meets standards. 1: Needs improvement/Below standards\\
4. Holistic Integration: Combine your dimensional scores into a single holistic score from 0--100, considering the relative importance of each meta-dimension for rubric effectiveness.\\
5. Synthesize Feedback: Based on your meta-dimensional assessment, identify the single greatest strength and the one or two most critical areas for improvement. Synthesize this into a concise, actionable feedback paragraph (2--3 sentences). The feedback should first acknowledge any strengths and then clearly state the areas needing improvement, referencing the relevant meta-rubric dimensions (e.g., Criteria Alignment, Level Distinction, Descriptive Language).\\
6. Quality Assurance: Ensure your score reflects the actual rubric quality demonstrated and your feedback is constructive and specific.}\\
\texttt{</instructions>}\\
\\
\texttt{<output format>}\\
\texttt{Return ONLY a JSON object in this exact format: \{ "score": [0--100], "comment": "[A concise paragraph...]" \}}\\
\texttt{</output format>}\\
\\
\texttt{\textbf{<USER PROMPT>}} \\
\texttt{<task description>}\\
\texttt{[-- A placeholder for the task description for the rubric --]}\\
\texttt{</task description>}\\
\\
\texttt{<rubric to evaluate>}\\
\texttt{[-- A placeholder for the user-designed rubric text --]}\\
\texttt{</rubric to evaluate>}\\
\\
\texttt{Please evaluate this rubric following the systematic meta-rubric methodology described above.}
\end{quote}

\subsubsection{Instruction-Based Revision Prompt}
\label{appendix:baseline-revision}
The following general-purpose revision prompt was used in the chatbox for both the \textbf{Text-LLM} and \textbf{Rubric-LLM} conditions across both experiments. It requires the user to manually formulate a specific instruction. Unlike iRULER's \textit{How To} feature, it is designed to return no explanatory rationale for its changes. 
\textit{Temperature: 0.2}.

\begin{quote}
\scriptsize
\setlength{\baselineskip}{6pt} 
\setlength{\parskip}{0pt} 
\texttt{\textbf{<USER PROMPT>}} \\
\texttt{<role, task>}\\
\texttt{You are an expert assistant. You will revise the <user artifact> according to the specific <user instruction>. Your goal is to make precise modifications that fulfill the user's request while preserving the overall quality and style of the artifact.}\\
\texttt{</role, task>}\\
\\
\texttt{<user artifact>}\\
\texttt{[-- A placeholder for the user's current artifact --]}\\
\texttt{</user artifact>}\\
\\
\texttt{<user instruction>}\\
\texttt{[-- A placeholder for the user's manually typed instruction --]}\\
\texttt{</user instruction>}\\
\\
\texttt{<instructions>}\\
\texttt{1. Carefully analyze the <user instruction> to understand exactly what modifications are requested.}\\
\texttt{2. Revise the <user artifact> to fulfill the user's instruction while:}\\
\texttt{   - Maintaining the overall structure and flow of the text}\\
\texttt{   - Preserving the original style and tone unless specifically asked to change it}\\
\texttt{   - Maintaining the original meaning}\\
\texttt{   - Ensuring grammatical correctness and coherence}\\
\texttt{   - Making ONLY the necessary changes to address the user's request}\\
\texttt{3. Return the revised artifact in the specified JSON format.}\\
\texttt{</instructions>}\\
\\
\texttt{<output format>}\\
\texttt{[-- A placeholder for the revised artifact --]}\\
\texttt{</output format>}
\end{quote}

\subsection{Prompts in Rubric Design Panel}
\label{appendix:rubric-design-prompts}



\subsubsection{Prompt for Recommending a Criterion: }
\label{appendix:recommend-criterion-prompt}
This prompt is used to generate a new, relevant criterion for the rubric, including a name and a full set of level descriptors.
\textit{Temperature: 0.8}.

\begin{quote}
\scriptsize
\setlength{\baselineskip}{6pt} 
\setlength{\parskip}{0pt} 
\texttt{\textbf{<USER PROMPT>}} \\
\texttt{<role>}\\
\texttt{You are a knowledgeable and professional rubric writing tutor who is proficient in making criteria for rubrics for writing tasks.}\\
\texttt{</role>}\\
\\
\texttt{<task>}\\
\texttt{You are tasked with making a criterion for a rubric for a writing task: [-- A placeholder for the task description --]. The rubric should have [-- number --] score levels, and the score levels are from score\_[-- number --] to score\_1. Higher scores reflect stricter requirements and better performance.}\\
\texttt{</task>}\\
\\
\texttt{<instructions>}\\
\texttt{[Step 1] Understand the writing task... The existing dimensions in the rubric are: [-- A comma-separated list of existing criteria --], the criterion you will make... should be different...}\\
\texttt{[Step 2] Make a criterion for the rubric. You should give a criterion name and the description across all score levels...}\\
\texttt{[Step 2.1] The criterion name should be a single word or short phrase that describes the criterion.}\\
\texttt{[Step 2.2] The criterion description should be a single string that describes the criterion. The description should be concise and clear.}\\
\texttt{[Step 3] Return a JSON object containing the criterion.}\\
\texttt{</instructions>}\\
\\
\texttt{<output format>}\\
\texttt{\{ "name": "...", "criteria": \{ "score\_3": \{ "text":...\}, ... \} \}}\\
\texttt{</output format>}
\end{quote}

\subsubsection{Prompt for Generating/Refining a Whole Criterion: }
\label{appendix:refine-criterion-prompt}
This prompt handles multiple operations for an entire criterion row, including generating descriptors from scratch (`Generate`) or refining existing ones (`Improve`, `Shorten`, `Elaborate`, `Bullet-point`). 
\textit{Temperature: 0.6}.

\begin{quote}
\scriptsize
\setlength{\baselineskip}{6pt} 
\setlength{\parskip}{0pt} 
\texttt{\textbf{<USER PROMPT>}} \\
\texttt{<role>}\\
\texttt{You are a knowledgeable and professional writing tutor who is proficient in building and improving criteria for rubrics for writing tasks.}\\
\texttt{</role>}\\
\\
\texttt{<task>}\\
\texttt{Given a criterion, shown in <criterion>, which may be incomplete, please [-- A placeholder for the operation type, e.g., "Improve" --] this criterion across all score levels for the [-- criterion name --] dimension of a rubric for the writing task: [-- task description --].The criterion has [-- score --] score levels, and the score levels are from score\_[-- number --] to score\_1. Higher scores reflect stricter requirements and better performance. The rubric should be in the format of <output format>.}\\
\texttt{</task>}\\
\\
\texttt{<criterion>}\\
\texttt{[-- A JSON object representing the criterion row --]}\\
\texttt{</criterion>}\\
\\
\texttt{<instructions>}\\
\texttt{Follow these steps to complete the task:}\\
\texttt{[Step 1] Understand the rubric dimension [-- criterion name --] and the provided <criterion> content based on your professional knowledge. '[-- description --]'}\\
\texttt{[Step 2] [-- refine type --] the [-- criterion name --] criterion across all score levels from 'score\_[-- score --]' to 'score\_1', ensuring consistency and logical progression where higher scores indicate better writing quality. Higher scores reflect stricter requirements and better performance.}\\
\texttt{[Step 3] Return only a JSON object with the updated criterion for all score levels, without any additional text or explanation. The output must match the <output format> and preserve the original text's case.}\\

\texttt{Note: The operation [-- refine type --] can be one of the following:
  - 'Generate': Create a new criterion from scratch.\\
  - 'Improve': Enhance clarity, detail, and overall quality.\\
  - 'Shorten': Make the descriptions more concise while retaining meaning.\\
  - 'Elaborate': Add more detail and depth to the descriptions.\\
  - 'Bullet-point': Format the descriptions as a single string with bullet points separated by line breaks (e.g., '• First point \\n• Second point').\\
  - 'Retry': Rewrite the criterion using a different approach.}\\
\texttt{</instructions>}\\
\\
\texttt{<output format>}\\
\texttt{\{ "score\_3": \{ "text": "...", "checked": false \}, ... \}}\\
\texttt{</output format>}
\end{quote}

\subsubsection{Prompt for Refining an Individual Descriptor: }
\label{appendix:refine-descriptor-prompt}
This prompt allows for more granular control, targeting a single performance level descriptor within a criterion. 
\textit{Temperature: 0.4}.

\begin{quote}
\scriptsize
\setlength{\baselineskip}{6pt} 
\setlength{\parskip}{0pt} 
\texttt{\textbf{<USER PROMPT>}} \\
\texttt{<role>}\\
\texttt{You are a knowledgeable and professional rubric writing tutor, you need to [-- refine type --] the description of a criterion for rubrics for writing tasks.}\\
\texttt{</role>}\\
\\
\texttt{<task>}\\
\texttt{You are tasked with [-- refine type --]ing the description of a criterion called [-- criterion name --]... The description is in the text field of the corresponding [-- score level --] object shown in <criterion>.}\\
\texttt{The criterion has [-- score --] score levels, and the score levels are from score\_[-- score --] to score\_1. Higher scores reflect stricter requirements and better performance.}\\
\texttt{The description is in the text field of the corresponding [-- score --] object shown in <criterion>.}\\
\texttt{The other scoring criteria can be either empty or filled in.}\\
\texttt{</task>}\\
\\
\texttt{<criterion>}\\
\texttt{[-- A JSON object representing the current state of the criterion row, with the target descriptor identified --]}\\
\texttt{</criterion>}\\
\\
\texttt{<instructions>}\\
\texttt{Follow these steps to complete the task:}\\
\texttt{[Step 1] Understand the rubric dimension [-- criterion name --] and the provided <criterion> content based on your professional knowledge. '[-- description --]'}\\
\texttt{[Step 2] [-- refine type --] the [-- criterion name --] criterion across all score levels from 'score\_[-- score --]' to 'score\_1', ensuring consistency and logical progression where higher scores indicate better writing quality. Higher scores reflect stricter requirements and better performance.}\\
\texttt{[Step 3] Return only a JSON object with the updated criterion for all score levels, without any additional text or explanation. The output must match the <output format> and preserve the original text's case.}\\

\texttt{Note: The operation [-- refine type --] can be one of the following:
  - 'Generate': Create a new criterion from scratch.\\
  - 'Improve': Enhance clarity, detail, and overall quality.\\
  - 'Shorten': Make the descriptions more concise while retaining meaning.\\
  - 'Elaborate': Add more detail and depth to the descriptions.\\
  - 'Bullet-point': Format the descriptions as a single string with bullet points separated by line breaks (e.g., '• First point \\n• Second point').\\
  - 'Retry': Rewrite the criterion using a different approach.}\\
\texttt{</instructions>}\\
\\
\texttt{<output format>}\\
\texttt{\{ "[-- score level --]": \{ "text": "...", "checked": false \} \}}\\
\texttt{</output format>}
\end{quote}

\subsubsection{Prompt for Enhancing the Task Description: }
\label{appendix:enhance-description-prompt}
This prompt is used to enrich a user's initial, often brief, task description into a more structured and comprehensive one. 
\textit{Temperature: 0.2}.

\begin{quote}
\scriptsize
\setlength{\baselineskip}{6pt} 
\setlength{\parskip}{0pt} 
\texttt{\textbf{<USER PROMPT>}} \\
\texttt{<role>}\\
\texttt{You are a professional rubric writing tutor, skilled in analyzing and concisely elaborating writing task descriptions for rubric creation.}\\
\texttt{</role>}\\
\\
\texttt{<task>}\\
\texttt{You are tasked with elaborating the writing task in [-- user's initial text --] into a concise, structured description with newline-separated bullet points, organized by thematic sections, while preserving the original intent.}\\
\texttt{</task>}\\
\\
\texttt{<instructions>}\\
\texttt{1. \textbf{Analyze the Task}: - Read the <user input> to identify the task type (e.g., essay, report), purpose (e.g., persuade, inform), and any missing details (e.g., audience, scope).
     - Make minimal, reasonable assumptions for missing details based on task context.}\\
\texttt{2. \textbf{Elaborate Concisely}: \\
        - Expand the <user input> into a concise description, one sentence per section, including the following sections:\\
        - Task Objective: State the main goal in one sentence.\\
        - Topic (If the task is given a topic): List the main topic of the task in one sentence.\\
        - Context: Describe the task's setting or purpose in one sentence.\\
        - Requirements: Length Requirements (e.g., 100 words).\\
        - Keep each bullet point short, focused, and non-redundant (max 20 words per point).\\
        - Use newline characters to separate sections for readability.}\\
\texttt{3. \textbf{Return Output}:  Return a JSON object with the elaborated description as a single string in the <output format> field.\\
     - Use clear section labels (e.g., "Task Objective, Topic, Context, Requirements").\\
     - Exclude any text outside the JSON structure.}\\
\texttt{</instructions>}\\
\\
\texttt{<output format>}\\
\texttt{\{ "task": "give the elaborated description as a single string with line breaks." \}}\\
\texttt{</output format>}
\end{quote}

\section{Rubric JSON Representation}
\label{appendix:rubric-json}

In \name{}, all rubrics are stored and exchanged in a unified JSON format. Each rubric consists of a list of criteria; each criterion has a name, a weight, and a set of level descriptors. 
Specifically, a rubric is represented as:

{
\scriptsize
\begin{verbatim}
{
  "rubric": [{
      "criteria_name": "<CRITERIA_NAME>",
      "percentage": <WEIGHT_IN_PERCENT>,
      "criteria": [{
          "score_1": {
            "text": "<LEVEL_1_DESCRIPTION>",
            "checked": <true_or_false>,
            "reason": "<MARKDOWN_REASON_FOR_SELECTING_OR_NOT_SELECTING_LEVEL_1>"
          },
          "score_2": {
            "text": "<LEVEL_2_DESCRIPTION>",
            "checked": <true_or_false>,
            "reason": "<MARKDOWN_REASON_FOR_SELECTING_OR_NOT_SELECTING_LEVEL_2>"
          },
          "score_3": {
            "text": "<LEVEL_3_DESCRIPTION>",
            "checked": <true_or_false>,
            "reason": "<MARKDOWN_REASON_FOR_SELECTING_OR_NOT_SELECTING_LEVEL_3>"
          },
          ...
        }]
    },
    ...
  ]
}
\end{verbatim}
}

\section{Additional Results}

\subsection{Justified and Actionable Feedback Usage in iRULER}

To understand how participants engaged with iRULER, we logged feature usage and helpfulness ratings in both experiments (Table~\ref{tab:feature-usage}). Across the Writing Revision and Rubric Creation experiments, participants most often opened \textit{Why Not} explanations to diagnose performance gaps (Writing: $M = 5.03$; Rubric: $M = 6.21$), while \textit{How To} examples were used more selectively (Writing: $M = 2.78$; Rubric: $M = 2.92$) but rated as highly helpful (Writing: $M = 2.56$, Rubric: $M = 2.33$ on a $-3$ to $+3$ scale). 


\begin{table}[t]
    \centering
    \small 
    \caption[]{Usage statistics and perceived helpfulness ratings for the core feedback features within the \name{} condition. Helpfulness was rated on a 7-point Likert scale ranging from -3 (Strongly Disagree) to +3 (Strongly Agree).}
    \label{tab:feature-usage}
    
    \begin{tabular*}{\columnwidth}{@{\extracolsep{\fill}} l c c @{}}
        \toprule
        \textbf{\name{} Feature} & \textbf{Usage (M, SD)} & \textbf{Helpfulness (M, SD)} \\
        \midrule
        \multicolumn{3}{l}{\textbf{Writing Revision Experiment (\textit{N} $=$ 16)}} \\
        \midrule
        \hspace{3mm} \textit{Why} Explanations      & 3.69 (1.32) & 1.88 (0.81) \\
        \hspace{3mm} \textit{Why Not} Explanations  & 5.03 (1.64) & 2.00 (0.82) \\
        \hspace{3mm} \textit{How To} Examples       & 2.78 (1.42) & 2.56 (0.51) \\
        \midrule
        \multicolumn{3}{l}{\textbf{Rubric Creation Experiment (\textit{N} $=$ 12)}} \\
        \midrule
        \hspace{3mm} \textit{Why} Explanations      & 3.33 (1.49) & 2.25 (0.62) \\
        \hspace{3mm} \textit{Why Not} Explanations  & 6.21 (2.47) & 2.67 (0.49) \\
        \hspace{3mm} \textit{How To} Examples       & 2.92 (1.47) & 2.33 (0.78) \\
        \bottomrule
    \end{tabular*}
\end{table}

\section{Baseline System Interfaces}
\label{appendix:baseline-interface}

\begin{figure*}[t]
  \centering
  \includegraphics[width = 0.84\linewidth]{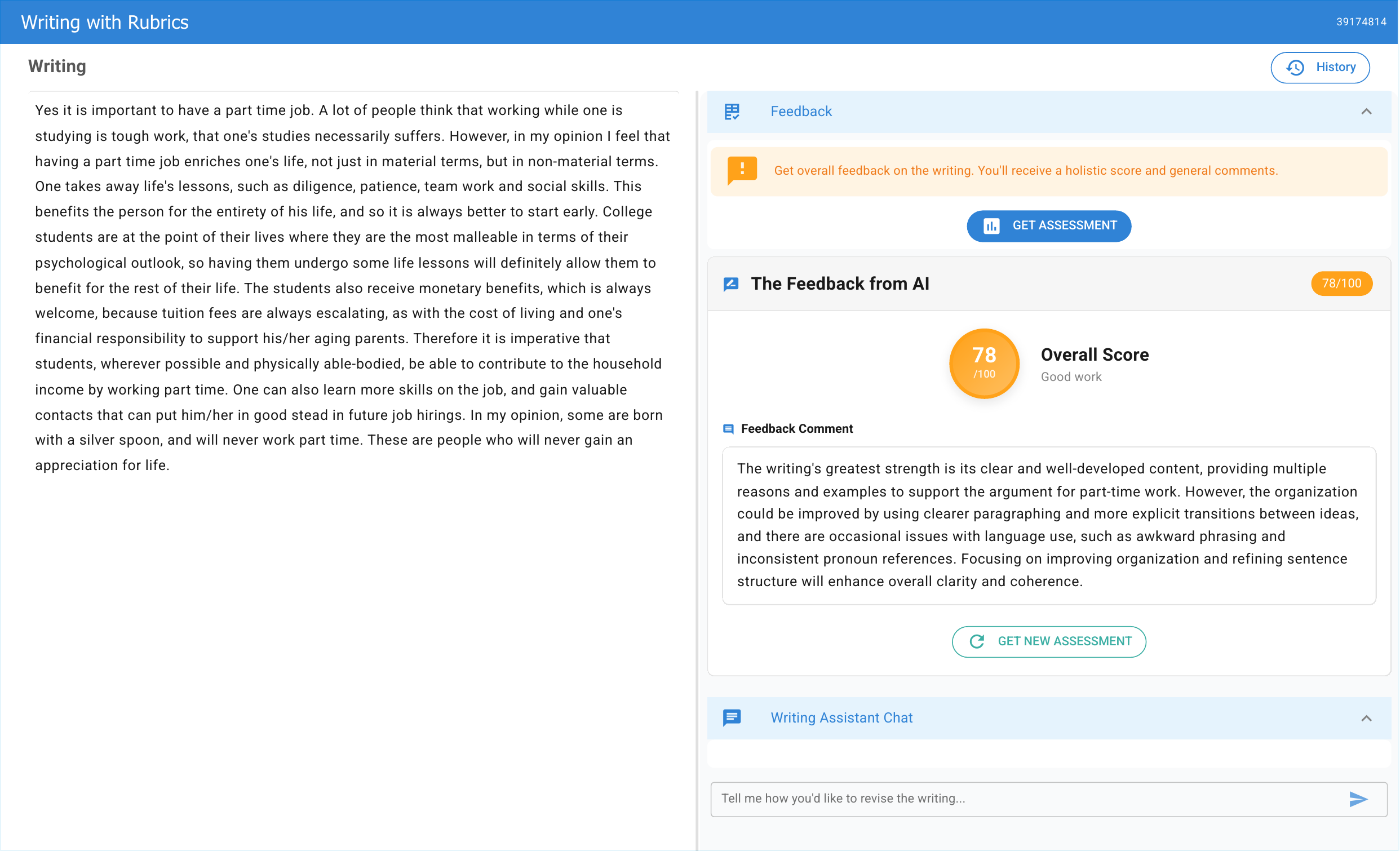}
  \caption[]{Screenshot of Writing Baseline system: Text-LLM}
  \Description{A screenshot of the "Writing Revision" user interface. The UI is divided into two vertical panels. The left panel, labeled "Writing," contains a multi-paragraph essay about the importance of part-time jobs for college students. The right panel, labeled "Feedback," is topped by a history button. Its main section, "The Feedback from AI," shows an overall score of 78 out of 100, labeled "Good work." Below the score, a "Feedback Comment" box contains a paragraph starting with "The writing's greatest strength is its clear and well-developed content..." and suggests improvement in paragraphing and language use. At the bottom are a "Get Assessment" button and an expandable text chatbox labeled "Writing Assistant Chat."}
  \label{fig:writing-baseline-text}
\end{figure*}

\begin{figure*}[t]
  \centering
  \includegraphics[width = 0.84\linewidth]{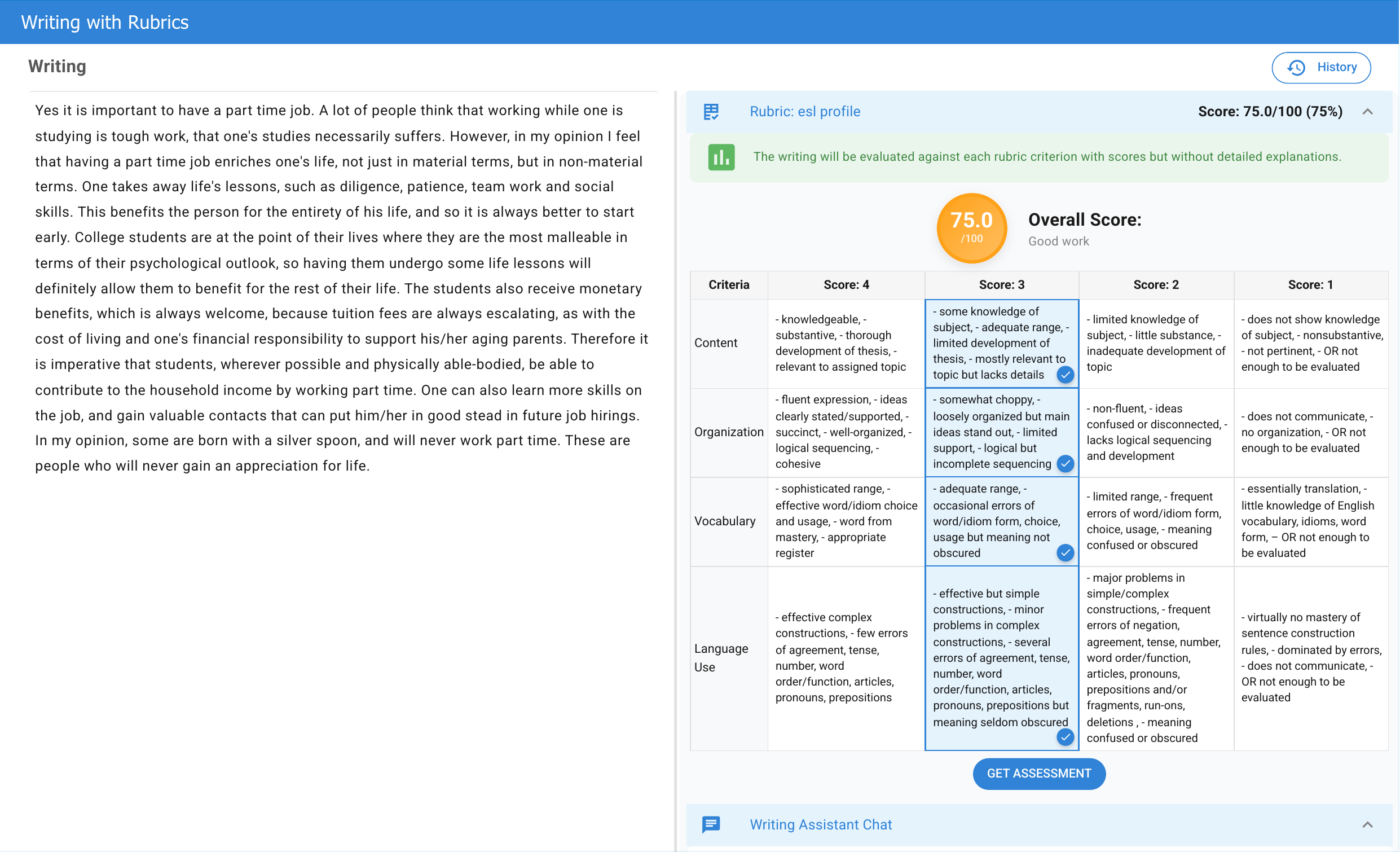}
  \caption[]{Screenshot of Writing Baseline system: Rubric-LLM}
  \Description{A screenshot of the "Writing Revision" user interface. The UI is divided into two vertical panels. The left panel, labeled "Writing," contains a multi-paragraph essay about the importance of part-time jobs for college students. The right panel, labeled "Feedback," is topped by a history button. Its main section, "The Feedback from AI," shows an overall score of 75.0 out of 100, labeled "Good work." Below is a detailed rubric table with four criteria rows: Content, Organization, Vocabulary and Language Use. Each row has four columns for scores from 1 to 4, with descriptive text in each cell. Checkmarks indicate the scores awarded for each criterion. At the bottom are a "Get Assessment" button and an expandable text chatbox labeled "Writing Assistant Chat."}
  \label{fig:writing-baseline-rubric}
\end{figure*}

\begin{figure*}[t]
  \centering
  \includegraphics[width = 0.84\linewidth]{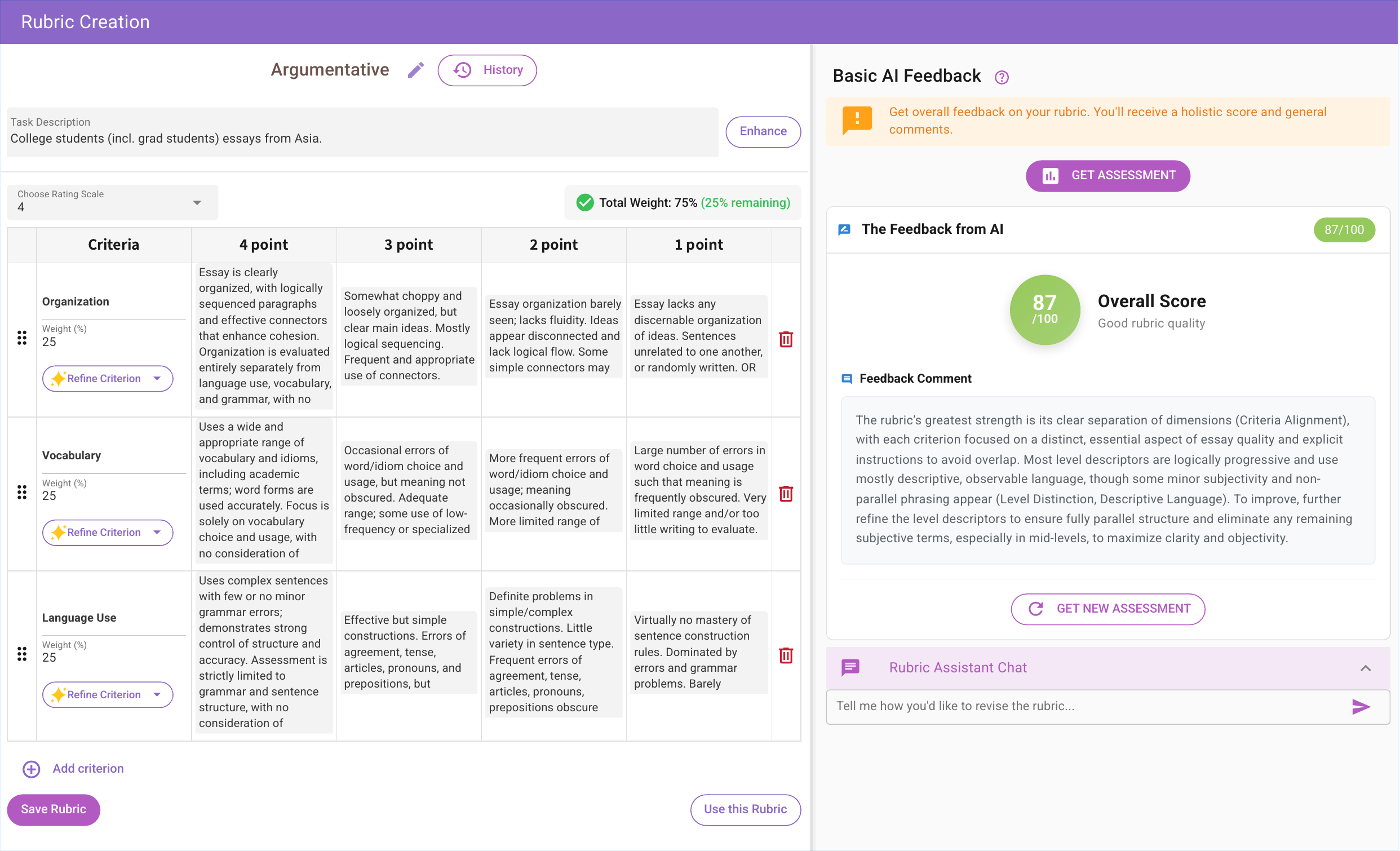}
  \caption[]{Screenshot of Rubric Baseline system: Text-LLM}
  \Description{A screenshot of the "Rubric Creation" user interface. The left panel shows a rubric being designed for an "Argumentative" essay, with criteria such as Organization, Vocabulary, and Language Use partially defined. The total weight is noted as 75 percent, with 25 percent remaining. The right panel, labeled "Basic AI Feedback," contains a section "The Feedback from AI" displaying an overall score of 87 out of 100. Below the score, a "Feedback Comment" box contains a paragraph starting with "The rubric's greatest strength is its clear separation of dimensions..." At the bottom are a "Get Assessment" button and an expandable text chatbox labeled "Rubric Assistant Chat."}
  \label{fig:rubric-baseline-text}
\end{figure*}

\begin{figure*}[t]
  \centering
  \includegraphics[width = 0.84\linewidth]{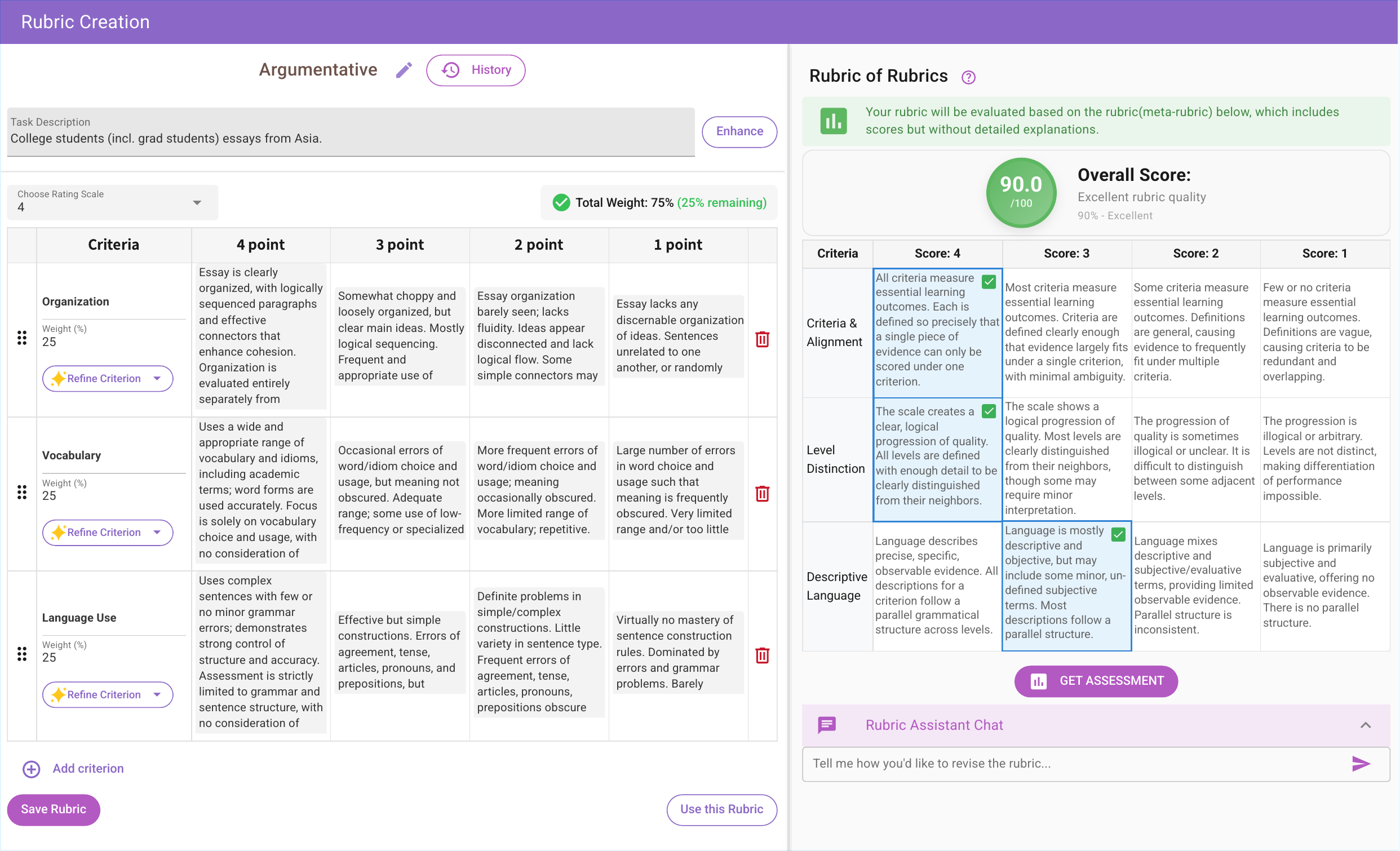}
  \caption[]{Screenshot of Rubric Baseline system: Rubric-LLM}
  \Description{A screenshot of the "Rubric Creation" user interface. The left panel shows a rubric under design for an "Argumentative" essay, with criteria including Organization and Vocabulary. The right panel, titled "Rubric of Rubrics," displays an overall score of 90.0 out of 100, labeled "Excellent." Below is a detailed meta-rubric table with three criteria rows: Criteria Alignment, Level Distinction, and Descriptive Language. Each row has four columns for scores from 1 to 4, with descriptive text in each cell. Checkmarks indicate the scores awarded for each meta-criterion. At the bottom are a "Get Assessment" button and an expandable text chatbox labeled "Rubric Assistant Chat."}
  \label{fig:rubric-baseline-rubric}
\end{figure*}

Fig. ~\ref{fig:writing-baseline-text}, \ref{fig:writing-baseline-rubric}, \ref{fig:rubric-baseline-text}, and~\ref{fig:rubric-baseline-rubric} illustrate the user interfaces of the baseline systems employed across our experimental conditions.



\section{Rubric Used in the User Study}
\label{appendix-user-study-rubric}

\begin{table*}[t]
  \centering
  \footnotesize
  \renewcommand{\arraystretch}{1}
  \begin{tabular}{p{0.06\linewidth}p{0.20\linewidth}p{0.23\linewidth}p{0.23\linewidth}p{0.19\linewidth}}
    \toprule
    \textbf{Criteria} & \textbf{Score 4} & \textbf{Score 3} & \textbf{Score 2} & \textbf{Score 1} \\
    \midrule
    Content (30\%)
    &
    knowledgeable; substantive; thorough development of thesis; relevant to assigned topic
    &
    some knowledge of subject; adequate range; limited development of thesis; mostly relevant to topic but lacks details
    &
    limited knowledge of subject; little substance; inadequate development of topic
    &
    does not show knowledge of subject; nonsubstantive; not pertinent; or not enough to be evaluated
    \\
    \midrule
    Organization \ (20\%)
    &
    fluent expression; ideas clearly stated/supported; succinct; well-organized; logical sequencing; cohesive
    &
    somewhat choppy; loosely organized but main ideas stand out; limited support; logical but incomplete sequencing
    &
    non-fluent; ideas confused or disconnected; lacks logical sequencing and development
    &
    does not communicate; no organization; or not enough to be evaluated
    \\
    \midrule
    Vocabulary \ (20\%)
    &
    sophisticated range; effective word/idiom choice and usage; word form mastery; appropriate register
    &
    adequate range; occasional errors of word/idiom form, choice, or usage but meaning not obscured
    &
    limited range; frequent errors of word/idiom form, choice, or usage; meaning confused or obscured
    &
    essentially translation; little knowledge of English vocabulary, idioms, or word forms; or not enough to be evaluated
    \\
    \midrule
    Language Use \ (30\%)
    &
    effective complex constructions; few errors of agreement, tense, number, word order/function, articles, pronouns, or prepositions
    &
    effective but simple constructions; minor problems in complex constructions; several errors of agreement, tense, number, word order/function, articles, pronouns, or prepositions but meaning seldom obscured
    &
    major problems in simple or complex constructions; frequent errors of negation, agreement, tense, number, word order/function, articles, pronouns, or prepositions and/or fragments, run-ons, deletions; meaning confused or obscured
    &
    virtually no mastery of sentence construction rules; dominated by errors; does not communicate; or not enough to be evaluated
    \\
    \bottomrule
  \end{tabular}
  \caption[]{Writing Revision Experiment rubric: ESL writing profile rubric excluding Mechanics criteria.}
  \label{tab:esl-profile-rubric}
\end{table*}

\begin{table*}[t]
  \centering
  \footnotesize
  \renewcommand{\arraystretch}{1}
  \begin{tabular}{p{0.06\linewidth}p{0.21\linewidth}p{0.234\linewidth}p{0.216\linewidth}p{0.19\linewidth}}
    \toprule
    \textbf{Criteria} & \textbf{Score 4} & \textbf{Score 3} & \textbf{Score 2} & \textbf{Score 1} \\
    \midrule
    Criteria Alignment \ (30\%)
    &
    All criteria measure essential learning outcomes. Each is defined so precisely that a single piece of evidence can only be scored under one criterion.
    &
    Most criteria measure essential learning outcomes. Criteria are defined clearly enough that evidence largely fits under a single criterion, with minimal ambiguity.
    &
    Some criteria measure essential learning outcomes. Definitions are general, causing evidence to frequently fit under multiple criteria.
    &
    Few or no criteria measure essential learning outcomes. Definitions are vague, causing criteria to be redundant and overlapping.
    \\
    \midrule
    Level Distinction \ (40\%)
    &
    The scale creates a clear, logical progression of quality. All levels are defined with enough detail to be clearly distinguished from their neighbors.
    &
    The scale shows a logical progression of quality. Most levels are clearly distinguished from their neighbors, though some may require minor interpretation.
    &
    The progression of quality is sometimes illogical or unclear. It is difficult to distinguish between some adjacent levels.
    &
    The progression is illogical or arbitrary. Levels are not distinct, making differentiation of performance impossible.
    \\
    \midrule
    Descriptive Language \ (30\%)
    &
    Language describes precise, specific, observable evidence. All descriptions for a criterion follow a parallel grammatical structure across levels.
    &
    Language is mostly descriptive and objective, but may include some minor, undefined subjective terms. Most descriptions follow a parallel structure.
    &
    Language mixes descriptive and subjective/evaluative terms, providing limited observable evidence. Parallel structure is inconsistent.
    &
    Language is primarily subjective and evaluative, offering no observable evidence. There is no parallel structure.
    \\
    \bottomrule
  \end{tabular}
  \caption[]{Rubric Creation Experiment rubric: Rubric-of-rubrics for evaluating rubric quality.}
  \label{tab:rubric-of-rubrics}
\end{table*}


\end{document}